\documentclass{article}
\usepackage[a4paper, margin=1in]{geometry}
\usepackage{graphicx}
\graphicspath{{figures/}}

\usepackage{abstract}

\usepackage{hyperref}

\usepackage{authblk}

\usepackage{float}
\floatstyle{plaintop}
\restylefloat{table}

\usepackage{amsmath}
\usepackage{amssymb} 

\newcommand*{\mysqrt}[4]{\sqrt[\leftroot{#1}\uproot{#2}#3]{#4}}

\usepackage{booktabs}

\usepackage{algorithm}
\usepackage{algorithmicx}
\usepackage{algpseudocode}

\usepackage{natbib}

\title{ZINBGT: Exploratory Data Analysis of Single-Cell Transcriptomic Expression Using Mixture Models}

\author[1]{Toby Kettlewell\footnote{Corresponding author: \href{email:Toby.Kettlewell@Glasgow.ac.uk}{Toby.Kettlewell@Glasgow.ac.uk}}}
\author[2]{Yiyi Cheng}
\author[2,3,4]{Thomas D. Otto}
\author[1]{\\Vincent Macaulay}
\author[1]{Mayetri Gupta}
\affil[1]{School of Mathematics and Statistics, University of Glasgow, University Place, G12 8QQ, Glasgow, United Kingdom}
\affil[2]{School of Infection and Immunity, University of Glasgow, University Place, G12 8QQ, Glasgow, United Kingdom}
\affil[3]{Bernhard Nocht Institute for Tropical Medicine, Hamburg, Germany}
\affil[4]{University of Hamburg, Hamburg, Germany}
\date{}

\begin{document}

\maketitle
\begin{abstract}
Single-cell transcriptomic data approximates the abundance of proteins at a high resolution, but its noisiness necessitates transformation by a pipeline of methods before analysis and inference. In the absence of robust validation of these pipelines and methods, it remains unclear how best to process any particular dataset. To compensate for this, popular visualisation methods, e.g., t-SNE and UMAP, are commonly used to produce descriptions of datasets. Such visualisations are incomplete and provide subjective descriptions of samples rather than statistically meaningful statements about technical noise or biology.

In this paper, we introduce the Zero-Inflated Negative-Binomial with Geometric Tail (ZINBGT), a mixture-model-based strategy for producing interpretable visualisations of each gene's expression across cells, along with diagnostic summaries that use Wasserstein distance to highlight outlier genes. These diagnostics are used to reveal an outlier gene within a \textit{T. brucei} sample. 
This method is applied to a human immune-cell dataset, highlighting the relationship between sparsity, mean, and spread across genes, as well as revealing an issue with the use of zero-inflated negative-binomial distributions to model single-cell RNA data. An investigation of simulated datasets intended to replicate the immune-cell data revealed discrepancies with the ground truth, establishing purposes for which these simulated datasets are unsuitable. Finally, we list a number of different domains to which this method can be applied.\\
\textbf{Availability:} The code used to run the analyses and produce the figures in this document will be made available on GitHub (\href{https://github.com/TobyK95/ZINBGT}{https://github.com/TobyK95/ZINBGT}). Before this material has been released, it can be provided upon reasonable request.\\
\end{abstract}
\section{Introduction}
\label{sec:intr}
The creation and analysis of transcriptomic data has improved our understanding of RNA and consequently protein production \citep{lowe_transcriptomics_2017, buccitelli_mrnas_2020}. The development of single-cell RNA sequencing (scRNA-seq) allows for the profiling of RNA on a cellular level, providing data on the differences between cells \citep{jovic_singlecell_2022}. This facilitated research into cellular heterogeneity within and between organisms, as well as elucidating the impact of illnesses and medical interventions on cell function.

The substantial technical noise found in scRNA-seq data necessitates a sequence of preprocessing transformations, generally performed through an analysis pipeline. While recommended pipelines exist  \citep[e.g.,][]{slovin_single-cell_2021, heumos_best_2023-1}, there is no single gold standard and no method to objectively assess the suitability of any particular pipeline for a given sample \citep{lahnemann_eleven_2020, stephenson_single-cell_2021}. Often, a user iteratively changes steps and parameters in their pipeline during the course of analysis according to whether the summaries and visualisations of the transformed data appear consistent with the user's expectations. This subjective approach can compromise the validity and objectivity of an analysis.

A popular way to compare the suitability of different pipelines for a particular sample is to apply them to a simulated dataset intended to resemble true data. Since the simulated dataset has a known ground truth, one can compare the pipelines to each other using Exploratory Data Analysis, based on summary statistics and visualisations, selecting the one returning conclusions consistent with the simulated ground truth \citep{cao_benchmark_2021, crowell_shaky_2023}.
While superficially sound, this approach requires the simulated data to be meaningfully similar to the true data.  

Principal Component Analysis (PCA) \citep{pearson_lines_1901}, t-SNE \citep{van_der_maaten_visualizing_2008} and UMAP \citep{healy_uniform_2024} are often used to visualise single-cell transcriptomic data. The standard paradigm when using these methods is to treat the data as a collection of cells, with each cell's RNA content across genes providing coordinates in ``gene space". These methods then find a lower-dimensional embedding of these cells, intended to preserve the cell-to-cell distances. While this provides clear summaries of cellular similarities and clustering structure, no visualisation can ever fully describe a dataset, as all visualisations necessarily contain considerably less information than the full data.

While PCA technically can preserve the entirety of the information within a sample, its results are almost always as complex as the original data, unless all the cells happen to lie on a lower-dimensional, linear subspace. Consequently, only two or three principal components are typically used for visualisations. UMAP and t-SNE tend to capture the relevant structures more efficiently than PCA, but they depend on hyperparameters chosen subjectively. While they produce intuitively interpretable figures, they do not furnish us with rigorous mathematical information; therefore direct, objective comparisons of samples cannot be made.

When modelling scRNA-seq data, the zero-inflated negative-binomial distribution is popularly used  
\citep[e.g,][]{clivio_detecting_2019, cui_comprehensive_2023}. This mixture distribution extends the negative-binomial distribution (a standard distribution for count data) by allowing for higher probabilities that a count value will be zero, reflecting the sparsity of scRNA-seq datasets.

In this paper we propose a method, the Zero-Inflated Negative-Binomial with Geometric Tail (ZINBGT), which produces independent summaries of each gene's expression across cells by fitting a mixture model to each gene and visualising the parameter estimates. It extends the zero-inflated negative-binomial model by allowing the probability of a zero to be lower than a negative-binomial distribution would assume, and introducing a component with a geometric distribution that captures additional kurtosis. The mixture components are simplified or omitted if doing so decreases the Bayesian Information Criterion (BIC). The Wasserstein distance \citep{kantorovich_mathematical_1960} is used to compare each gene's fitted model to the data's empirical distribution. A bootstrap based analogue of a $p$-value, called a $p_B$-value, is introduced to assess whether a Wasserstein distance is consistent with adequate model fit.

The layout of the paper is as follows. Section \ref{sec:meth} will describe the ZINBGT model in detail, then define the Wasserstein-based diagnostics. Section \ref{sec:resu} will introduce and analyse a number of datasets, including a human immune-cell dataset and simulations designed to resemble it. Additionally, a \textit{Trypanosoma brucei} sample will be explored to reveal possible outliers. The implications of these visualisations will be explored in Section \ref{sec:conc}, and finally we will outline some areas to which ZINBGT can be applied.

\section{Methods}
\label{sec:meth}
ZINBGT provides concise summary statistics for each gene independently, by modelling each gene's RNA count across cells as a random variable generated from a mixture distribution. %
Firstly, we parametrise the negative-binomial component in terms of its mean and dispersion (variance divided by mean), labelled $m$ and $d$ respectively. The Poisson distribution is recovered when $d=1$.
Secondly, while zeroes are typically split between the negative-binomial component and constant zero component, we assign all zeroes to the constant-zero component and restrict the negative-binomial component to strictly positive values, i.e., we use a hurdle negative-binomial distribution. Note that $m$ and $d$ are defined according to the mean and variance of the equivalent non-hurdle distribution, therefore, when $m=0$, the hurdle negative-binomial distribution becomes a constant-one distribution.

Finally, the mixture model contains a geometric distribution as its third component, parametrised by its mean, $\mu_g$. For consistency with the negative-binomial component, this geometric component is only defined on non-zero count values, and its mean parameter is defined with respect to its non-hurdle equivalent. The probabilities of a count value belonging to each of the three components are denoted by $p_0$, $p_1$, and $p_2$, for the constant-zero, hurdle negative-binomial, and geometric component, respectively.

\subsection{Our model}

A ZINBGT random variable $X$ with parameters $\boldsymbol{\theta}=(p_0, p_1, p_2, m, d, \mu_g)$ has the mass function 
\begin{align}
    \nonumber
    \mathbb{P}(X=x|\boldsymbol{\theta}) = \ 
        &\ p_0\mathbb{I}[x=0]\\
        +  &\ p_1\text{NB}_h \left(x;\, r={m}/{(d-1)},\ p=d^{-1}\right )
        \nonumber \\ 
        + &\ p_2\text{Geom}_h\left(x;\, p=({1+\mu_g})^{-1}\right),
\end{align}
\noindent
for $x\in \mathbb N$, where $\mathbb I$ is the indicator function, which maps true statements to one and false to zero; $\text{NB}_h$ is the mass function of a hurdle negative-binomial random variable; and $\text{Geom}_h$ is the mass function of a hurdle geometric random variable. The ranges for the parameters are: $p_0, p_1,p_2 \in [0,1]; p_0+p_1+p_2=1$; $m \ge 0$; $d \ge 1$ and $\mu_g \ge 0$. Fully expanded versions of these mass functions can be seen in Appendix \ref{app:param_est} of the Supplementary materials (eqns.\ (\ref{eqn:f0_def}), (\ref{eqn:f1_def}) and (\ref{eqn:f2_def})). Note that this model can only be applied to count data, because it has been defined using discrete distribution functions.

This model can be simplified in a number of ways; the negative-binomial component has as limiting cases a Poisson distribution ($d=1$) and a constant one ($m=0$), and one or both of the negative-binomial and geometric components can be omitted. These simplifications are equivalent to restrictions on parameter values, the quantities we will be visualising, and can necessitate restricting other parameter values to maintain the identifiability of the model. For example, omitting the geometric component is equivalent to setting $p_2=0$, in which case all $\mu_g$ values are equivalent, so we impose $\mu_g=0$. Table \ref{table:comp_simps} lists the component simplifications used, and the associated parameter restrictions. 

When choosing how components should be simplified, we make use of the fact that the set of negative-binomial distributions contains the set of geometric distributions (i.e., negative-binomial distributions with $m=d-1$). Hence, omitting the negative-binomial component while retaining the geometric component -- modelling a gene with a zero-inflated geometric distribution -- is equivalent to restricting the negative-binomial parameter values to the geometric case ($m=d-1$) while dropping the geometric component ($p_2=0$). Although the two approaches produce the same model, the fitted parameter values differ. We found that the zero-inflated geometric model was typically assigned to lowly-expressed genes. As we intend for the geometric component to describe larger tail values and the negative-binomial smaller values, these genes are interpreted more consistently under the negative-binomial parametrisation.

\begin{table}[t]
\begin{center}
\begin{tabular}{ l l l }
Component & Change & Parameter restrictions \\ 
\hline
\hline
\textbf{Negative-binomial} & Poisson    & $d=1$\\   
 & Constant 1 & $m=0,\ d=1$    \\
 & Omitted & $p_2=0,\ m=d-1,\ \mu_g=0$    \\
\hline
 \textbf{Geometric} & Omitted & $p_2=0,\ \mu_g=0$
\end{tabular}
\end{center}
\caption{Table of component simplifications. In each row, the first column states which component is simplified; the second column states either which distribution that component adopts, or that it is omitted. The third column describes the restrictions each simplification places on the parameter estimates.
}
\label{table:comp_simps}
\end{table}
The submodels are formed by simplifying or omitting one or both of the non-zero components. The only submodel not included results from simplifying the negative-binomial component as a constant one and keeping the geometric component. This model, a zero-inflated, one-inflated, geometric distribution, is not believed to meaningfully describe scRNA-seq data, and it is not worth the increased computational cost. The negative-binomial component can become a constant one when the geometric component is omitted.

Parameter estimation is performed for all submodels over all genes, except when all observations are zero or one, in which case model selection and parameter estimation are trivial. Maximum-likelihood (ML) estimation, either directly or using the EM algorithm, \citep{dempster_maximum_1977} is used for all submodels, with the parameter estimate being selected to minimise the BIC. This ensures complex models are only used when they substantially improve model fit.
For the submodels where one or both non-zero components have been omitted, ML estimates of the parameters can be calculated directly. Otherwise, the ML estimates are approximated using the EM algorithm.
This algorithm makes use of each observation's probability of belonging to each component as auxiliary variables. 
It alternates between estimating the parameter values while treating the component membership probabilities as known, and updating the component membership probabilities using the parameter estimates. Appendix \ref{app:param_est} provides the full details on model fitting.

While the EM algorithm always approaches a local optimum, we cannot be assured that the global optimum has been found. This is often mitigated by applying the algorithm multiple times with differing initial values. The increase in computation time would be excessive for a visualisation algorithm, without providing assurance of a global maximum. 
We therefore implement an initialisation strategy intended to be more likely to return parameter estimates with lower BIC values.
A number of initialisation strategies were experimented with to establish which performed best (see Appendix \ref{app:init_strat} in the Supplement for details).

\subsection{Wasserstein distance}

After computing parameter estimates for a gene,  metrics for the quality of fit are generated using the Wasserstein distance \citep{kantorovich_mathematical_1960}. The Wasserstein distance, for a cost parameter $\alpha\ge 1$, is a function that maps pairs of probability measures to non-negative numbers, with smaller values reflecting a greater similarity between distributions. A common analogy treats the two measures as different ways of piling a mass of earth, with the Wasserstein distance being the minimum cost of moving the earth from one configuration into the other. The cost of moving a section of earth increases with the distance travelled, with larger values of $\alpha$ increasing the impact of distance on cost. The Wasserstein distance between a pair of probability measures $\mu$ and $\eta$ (with the natural numbers as supports) is defined as 
\begin{equation}
W_\alpha(\mu, \eta)  = 
    \min_{\gamma \in \Gamma(\mu,\eta)}
    \mysqrt{0}{0}{\scriptstyle \alpha}{
            \sideset{}{_{x=0}^{\infty}}\sum
            \sideset{}{_{y=0}^{\infty}}\sum
                \gamma(x,y) |x-y|^{\alpha} }
\end{equation}
where $\Gamma(\mu, \eta)$ is the set of all joint measures with marginal measures $\mu$ and $\eta$.

The Wasserstein distance between the fitted model probability function and the empirical distribution of count values acts as a proxy for how well the model describes the data.  Changing the cost parameter, $\alpha$,  changes which types of differences are more strongly penalised; this can also be controlled by transforming the count values before calculating the Wasserstein distance. We compared settings for this penalty and transformation, from which it was concluded that a cost parameter of 1 and the transformation $x \mapsto \log(1+x)$, commonly called \texttt{log1p} in a number of programming languages, was better for identifying appropriate model fit (see Appendix \ref{app:wass_experiments} for details).

While the Wasserstein distance quantifies the differences between data and a model fitted to that data, by itself it does not tell us whether the difference is consistent with a fitted model describing the data well. 
To overcome this limitation, we used bootstrapping.  Once a model was fitted and the Wasserstein distance calculated, multiple bootstrap samples with the same sample size as the original data were generated from the fitted model. 
Because these sample distributions resemble the fitted model, while being the same size as the data, they should be representative of the kind of datasets we would have seen if model fitting was successful; and their Wasserstein distances representative of what we should expect if the parameter estimates are reflective of the real data.
We calculated the proportion of bootstrap Wasserstein distances larger than the distance between the fitted model and the true data, which we call the $p_B$-value and is analogous to the $p$-value used in hypothesis testing. Genes were sorted by their $p_B$-values, allowing a user to quickly identify genes for which the ZINBGT model is not suitable or parameter estimation was unsuccessful. Supplementary Algorithm \ref{alg:wass_boot} outlines the calculation of $p_B$-values, and Appendix \ref{appsub:pb_isnt_p} elaborates on the differences between $p_B$-values and $p$-values. 

\subsection{Speed}
Computation time is important for any visualisation method. Because ZINBGT performs parameter estimation and diagnostic value calculation separately for each gene, it is highly amenable to parallelisation. Calculation time increases linearly with the number of genes, and similarly it decreases linearly with the number of cores used.

Within a gene, computations use the unique RNA counts and the number of cells with that count. As such, computation time does not scale with the number of cells, but the number of unique count values for each gene. This means that computation time scales more slowly than linearly with the cell count. This is particularly relevant for the lowly expressed genes, for which there are few unique values. If a gene has no count values larger than one, then the usual calculations are skipped, as only the proportion of zero counts ($p_0$) needs to be computed. In this case, Wasserstein-based diagnostics are also omitted, because the model is trivially appropriate.

Calculating Wasserstein diagnostics, particularly $p_B$-values, is slower than parameter estimation. Users who needs to quickly iterate on processing a dataset can omit Wasserstein diagnostics entirely, or just calculate the Wasserstein distances. Alternatively, they can speed up the calculation of Wasserstein-based diagnostics by only investigating genes with large enough count values to be plausibly problematic.

\section{Results}
\label{sec:resu}

We centred our analysis on a subset of the immune-cell data from \citet{stephenson_single-cell_2021}. First, we explored the raw CD14 monocyte data, 
then we compared two simulations generated with the intention of mimicking this dataset. These were generated using \textit{Hierarchicell} \citep{zimmerman_hierarchicell_2021} and \textit{muscat} \citep{crowell_muscat_2020}. See Appendix \ref{appsub:sim_alg} for details.
Finally, we applied ZINBGT to a \textit{T. brucei} dataset \citep{briggs_single-cell_2021}.
The CD14 dataset was produced by downloading the h5ad files provided in \citet{stephenson_single-cell_2021}, and selecting the cells assigned to CD14 in the $\texttt{initial\_clustering}$ metadata. In order ensure the CD14 dataset was not too large to feasibly work with, only cells belonging to patients assigned a worst clinical status of ``Healthy" and ``Moderate" were kept. The \textit{T. brucei} dataset is the union of the three CSV files found on the Zenodo page linked by \citet{briggs_single-cell_2021}. Certain genes were found in some, but not all, of the files; these were removed from the analysis.
Computation times for these analyses can be seen in Appendix \ref{app:time_table}, along with each sample's gene and cell count. 

\subsection{CD14 monocytes}
\label{subsec:res_cd14}
\begin{figure}[t]
\centering
{\includegraphics{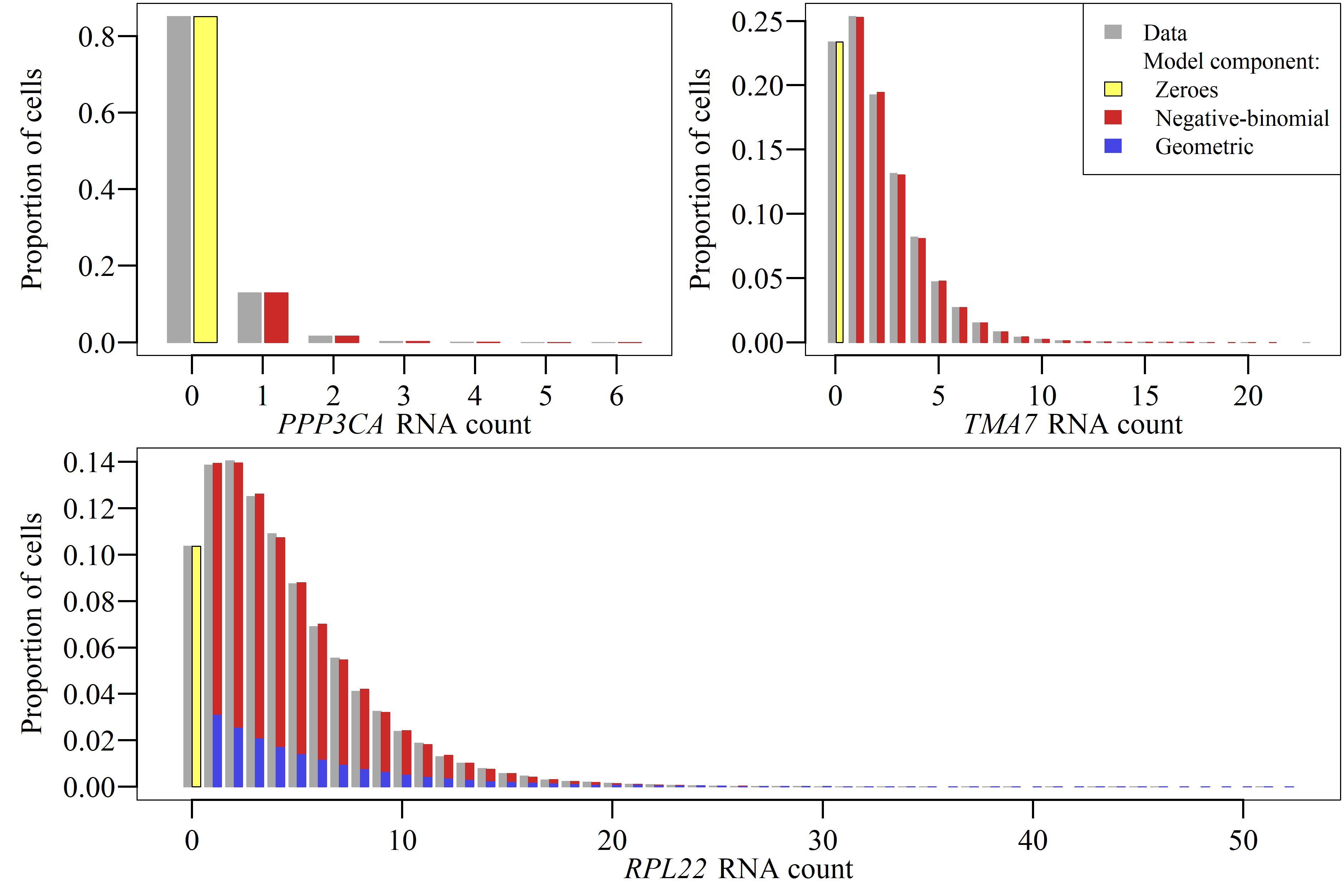}}
\caption{Bar charts showing the empirical and model mass functions of RNA counts in the CD14 sample for genes: \textit{PPP3CA}, \textit{TMA7} and \textit{RPL22}. 
}
\label{fig:cd1f_pmfs}
\end{figure}

Figure \ref{fig:cd1f_pmfs} gives three examples of observed gene expression in the CD14 sample alongside their fitted distributions. For the first two genes, the geometric component was dropped, as the negative-binomial component captured both the majority of non-zero values and the larger tail values. The third gene retained the geometric component, as it was necessary to increase the fitted distribution's kurtosis to more closely resemble the data.

\begin{figure}[t]%
\centering
{\includegraphics{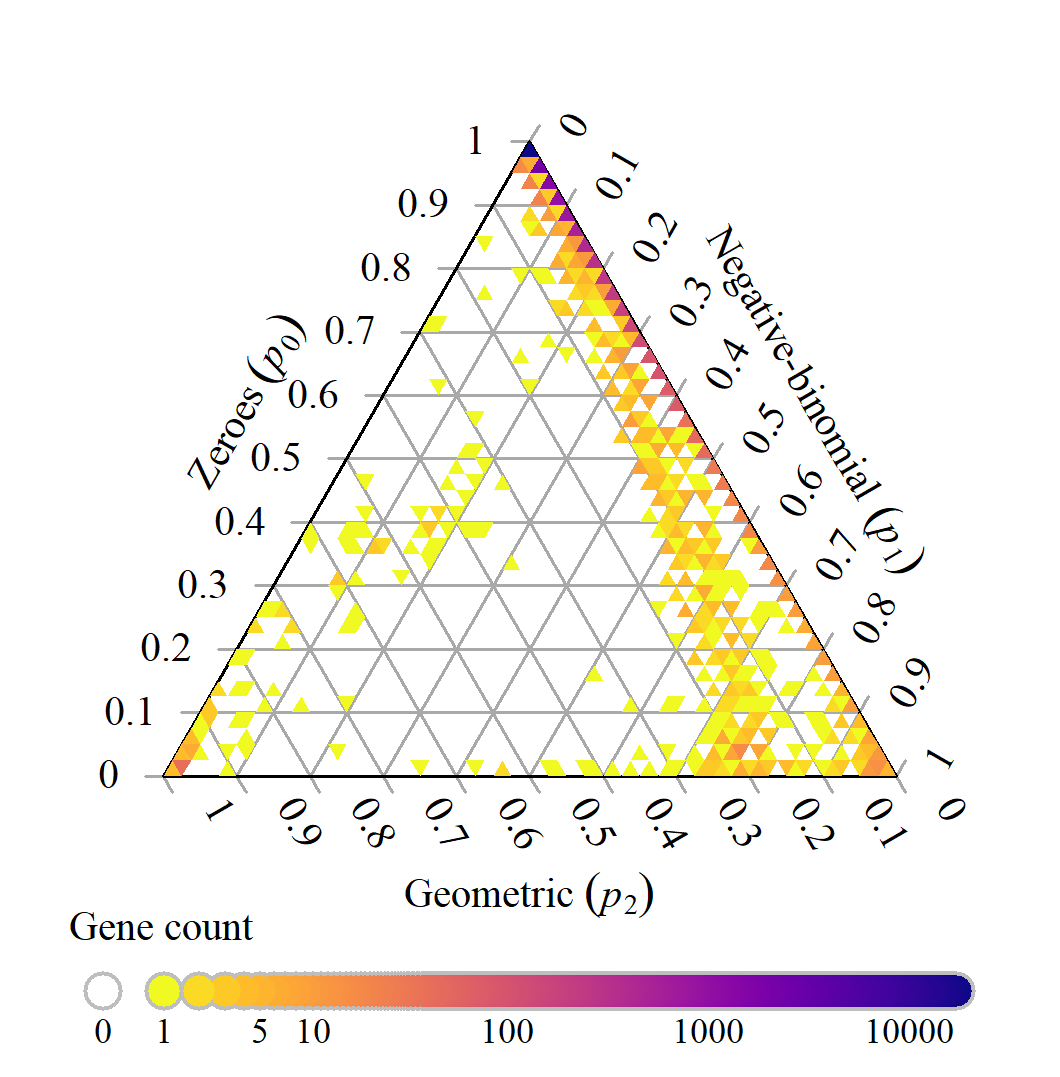}}
\caption{Ternary plot of the estimates of $(p_0, p_1, p_2)$ for each gene in the CD14 monocyte sample. Each small triangle represents a range of $p_0$, $p_1$ and consequently $p_2$ values, with its colour representing the number of genes with parameter estimates within that range. }
\label{fig:cd14_tern}
\end{figure}

Figure \ref{fig:cd14_tern} shows the relationships between component proportions across genes. Most genes had large constant-zero components after model fitting, reflecting the data's sparsity. In particular, the mode is near $p_0=1$, where no RNA belonging to a particular gene has been found. Many genes were fitted with a geometric component, and the curved ridge on the ternary plot reflects a tendency for the geometric component to be more prominent relative to the negative-binomial component as the sparsity decreases. This showed that the zero-inflated negative-binomial distribution cannot accurately describe gene expression for highly-expressed genes.

The impact of sparsity on the other parameters summarising gene expression is captured in Figure \ref{fig:cd14_hists}, with a full set of bivariate and univariate histograms presented in Supplementary Figure \ref{fig:cd14_full_hists}. By comparing the mean of the negative-binomial component, $m$, against the sparsity $p_0$, two clusters of genes were observable. The group with very low $m$ values usually had high $p_0$ values, and were made up almost entirely of zeroes, with the few non-zero values being very small, often smaller than five. 
Due to their sparsity, it may be advisable to consider these genes as not being meaningfully expressed in this sample, and remove them as part of quality control.

For the remaining genes, $p_0$, sparsity, was negatively correlated with $d$ and $\mu_g$, the negative-binomial component's dispersion and the geometric component's mean. This highlighted a tendency for less sparse genes to have their variances grow faster than their means, and for their kurtosis to become larger than a negative-binomial distribution alone could explain. The former observation justifies models using negative-binomial distributions over Poisson distributions in transcriptomics, while the latter reinforces that the negative-binomial distribution alone is not able to describe the range of counts observed in highly-expressed genes.

\begin{figure*}[t]%
\centering
{\includegraphics{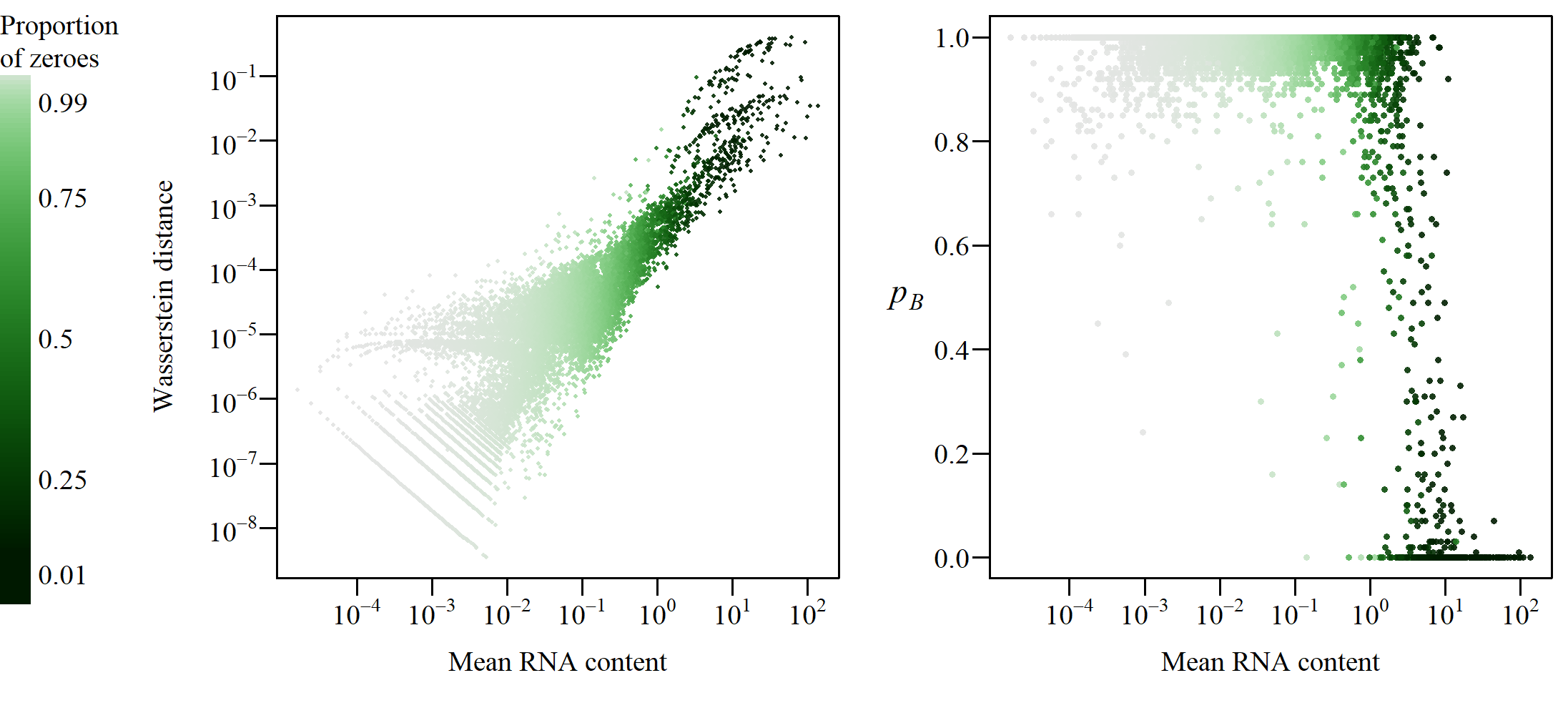}}
\caption[]{Scatter plots of each gene's mean RNA count with the Wasserstein distance between fitted model and empirical mass function (panel 1), and $p_B$-value (panel 2). Each striation in the bottom left of panel one consists of genes with the same, small number of two counts, with the other observations being zeroes and ones. Within a striation, genes differ in the proportion zeroes, hence their differing positions on the line. The horizontal mass of genes with a Wasserstein distance of approximately $10^{-5}$ are those with a single RNA count of three, with the other counts being lower. The remaining genes are dispersed across the figure continuously.
}
\label{fig:cd14_wass}

\end{figure*}

Figure \ref{fig:cd14_wass} provides scatter plots comparing each gene's RNA content against its Wasserstein distance and $p_B$-value.
It shows that a gene's mean RNA content correlated strongly with its Wasserstein distance, making it difficult to use Wasserstein distance to separate problematic from non-problematic genes. We also observed that mean RNA content was not as strongly correlated with $p_B$-values, making them more suitable for detecting problematic genes. As $p_B$-values are intended to streamline identifying the genes most likely to be problematic, we focussed on genes with $p_B$-values of zero (see Supplementary Figure \ref{fig:cd14_lowpb} for histograms describing some of these genes). Among these genes, those with the highest Wasserstein distance were discovered to have a kurtosis too large to be captured by the geometric component. The remaining genes' model distributions differed too little from the data to suggest that the parameter estimates were misleading. 
We therefore concluded that our visualisations of the fitted parameter values were highly representative of the underlying data for all but a few genes. 
This shows that our method was able to efficiently and intuitively summarise scRNA-seq datasets.

\begin{figure*}[t]%
\centering
{\includegraphics{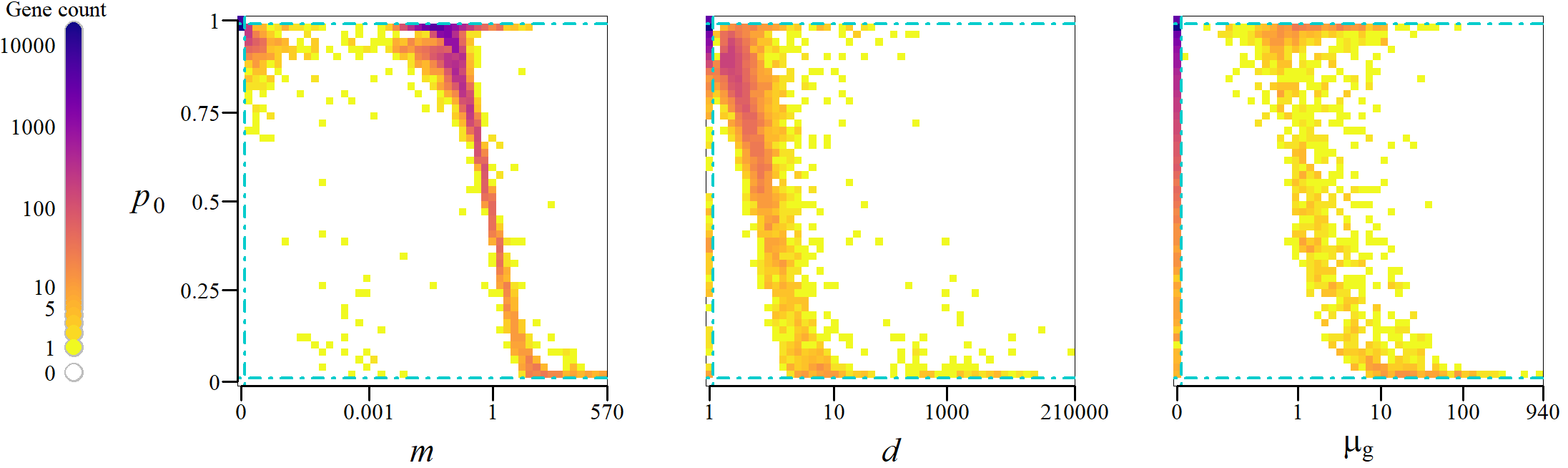}}
\caption{Two-dimensional histograms of $p_0$ against $m$, $d$, and $\mu_g$ for the genes in the CD14 dataset. Cyan dot-dashed lines separate boundary values from non-boundary values. E.g., the genes above the top horizontal lines are those for which $p_0$ is 1, whereas those just below have a $p_0$ value close to, but strictly smaller than, 1.}
\label{fig:cd14_hists}
\end{figure*}

\subsection{CD14 and its simulations}

\begin{figure}[t]%
\centering
{\includegraphics{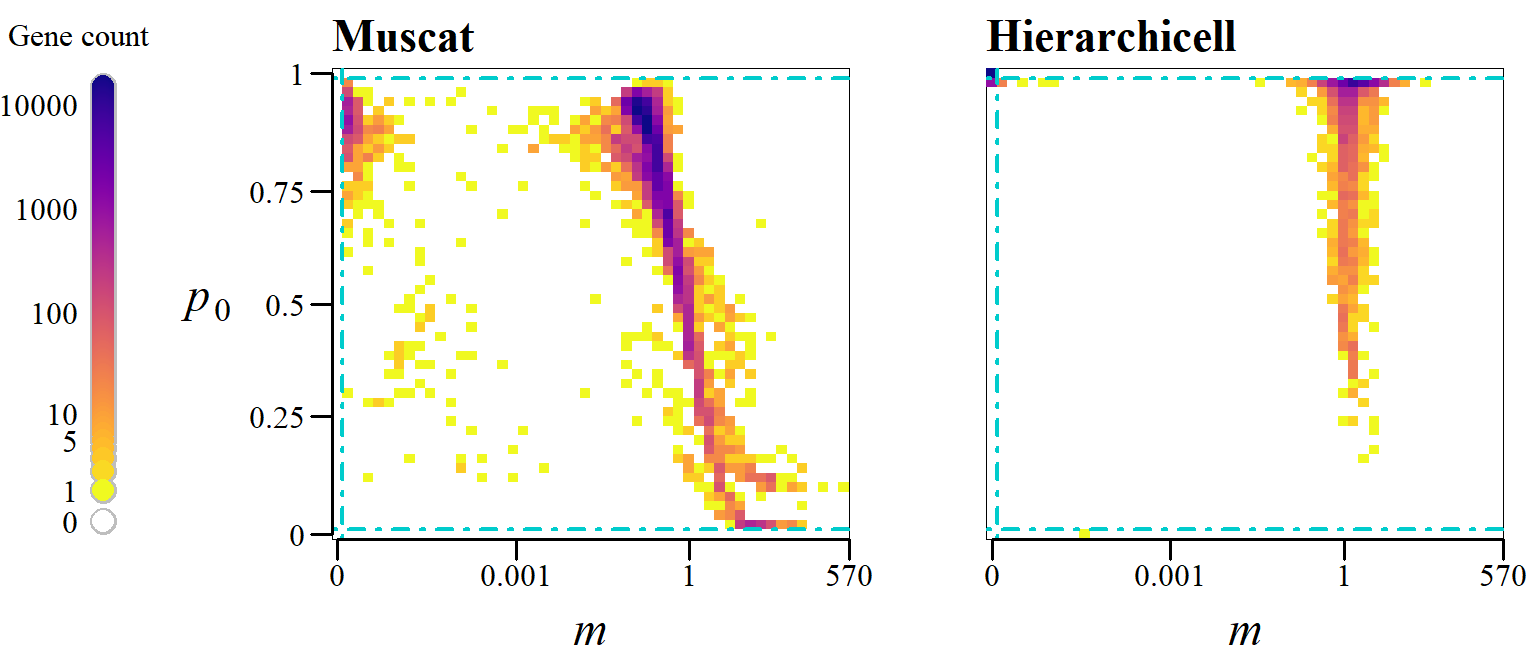}}
\caption{Two-dimensional histograms comparing $p_0$ and $m$ for genes in the samples simulated by \textit{muscat} and \textit{Hierarchicell} intended to replicate the real CD14 data.
}
\label{fig:cd14_sim_hists}
\end{figure}

We next used visualisations comparing the CD14 data to two simulated datasets that were intended to resemble it. In so far as a simulated dataset captures the properties of a real dataset, the summaries of simulated and real data should be similar. Our goal was to identify discrepancies between visualisations of real and simulated data that reveal a failure of the simulated data to reproduce the properties of the real data. The following analysis  therefore demonstrates the ability of ZINBGT visualisations to highlight the inadequacies of simulated datasets, which has implications regarding their usefulness in model validation.

The relationship between $p_0$ and $m$ seen in Figure \ref{fig:cd14_sim_hists} highlighted how each simulated sample captured some, but not all, of the characteristics of the true CD14 data seen in Figure \ref{fig:cd14_hists}. The \textit{muscat} generated sample showed a similar relationship between the sparsity and the mean of the negative component, although the same spread of $m$ values for genes with $p_0$ near 1 could not be seen. 
Similarly, the real data histograms contained  two dark pixels in the top left, representing the majority of genes for which the counts were almost all zeroes with few or no ones, but no such genes were found in the \textit{muscat} histogram.
The \textit{Hierarchicell} sample did contain these highly sparse genes, but it neither captured the relationship between $m$ and $p_0$, nor contained any genes fitted with large $m$ values, suggesting that this sample did not contain highly-expressed genes. The salient differences between the real CD14 sample and its two simulations confirmed that the simulated datasets were conspicuously unlike the real data, and reveal in part how they diverge.

\subsection{Outlier gene}
\label{subsec:resu_tbruce}
\begin{figure}[!t]%
\centering
{\includegraphics{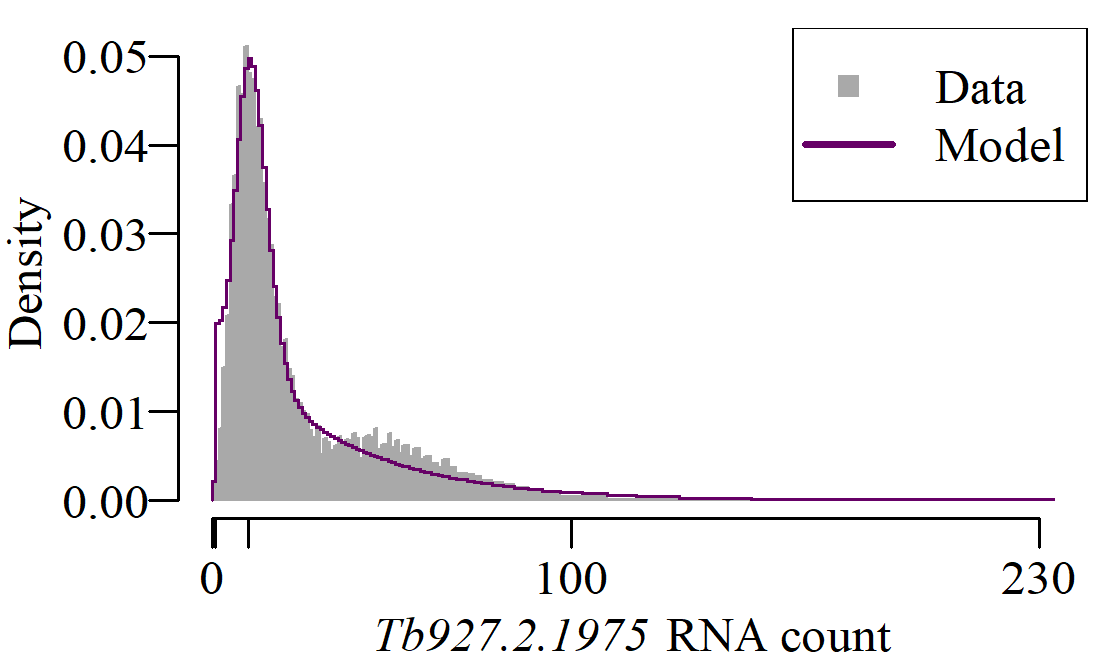}}
\caption{Histogram of count values observed for the outlier gene in a \textit{T. brucei} sample, \textit{Tb927.2.1975}
}\label{fig:t_bruce_bimod}
\end{figure}

When applying the ZINBGT method to the \cite{briggs_single-cell_2021} \textit{T. brucei} data, we found that almost all genes had low Wasserstein distances and high $p_B$-values consistent with the model being suitable and appropriately fitted (see Supplementary Figure \ref{fig:tbruce_wass} for a scatter plot of $p_B$-values against Wasserstein distances). Three genes differed from the others by having $p_B$-values of zero: \textit{Tb927.2.1975}, \textit{Tb08.27P2.260} and \textit{Tb927.1.4650}. Figure \ref{fig:t_bruce_bimod} shows the histogram of the gene with the highest Wasserstein distance (\textit{Tb927.2.1975}). While parameter estimation seemed to succeed in producing the ZINBGT distribution most similar to the observed data, it was incapable of exhibiting the appropriate bimodality, which was highlighted by our diagnostic values. Supplementary Figure \ref{fig:tbruce_pb_zero} compares the count data and fitted model distributions across the full sample and three subsamples for the three genes with $p_B$-values of zero. This revealed that the bimodality observed in gene \textit{Tb927.2.1975} was due to its distribution of RNA across cells differing substantially across the three subsamples. The gene with the next highest Wasserstein distance (\textit{Tb08.27P2.260}) had a model distribution that looked ambiguously bimodal, but comparing its distribution across the subsamples revealed a similar heterogeneity. Further exploration of the data highlighted a minority of genes with distributions that visibly differed across subsamples. Despite this, almost all genes were well described by the ZINBGT model, regardless of their underlying heterogeneity, including gene \textit{Tb927.1.4650}. This reinforced the ability of our method to describe scRNA-seq data, as well as showing that it was still useful when a minority of genes cannot be described. Here the applicability of our method to all but one gene showed that the gene is an outlier, and highlighted the power of the Wasserstein diagnostics in detecting such outliers.

\section{Discussion}
\label{sec:conc}
In this paper, we have introduced a novel mixture-model based visualisation strategy, expanding Exploratory Data Analysis by using parameter estimates as summary statistics. This allows for improved description and comparison of scRNA-seq data, critique of simulated datasets, and can refine our understanding of the distribution of RNA within cells.

\subsection{CD14 data}

Visualisations of the CD14 data make clear the massive sparsity of expression, with the bimodality in the 2D histograms highlighting genes that are not substantially expressed within a sample. This can be used to assess the effectiveness of quality control methods in removing genes for which there is little information. 
More striking was the inclusion of the geometric component for many genes, despite penalising model complexity. This showed that models using zero-inflated negative-binomial distributions may not adequately account for kurtosis. While including the geometric component often improved model fit, as quantified by Wasserstein diagnostics, the genes that were not well described were found to have too great a kurtosis for a geometric component to capture (see Supplementary Figure \ref{fig:cd14_lowpb}). This reinforced that models aiming to describe single-cell transcriptomic data need to account for its kurtosis, in addition to the standard considerations of sparsity, mean and variance.

\subsection{Simulations}

The genes in the true CD14 data can be split broadly into two categories: practically absent and not practically absent. While the \textit{Hierarchicell} sample contained an appropriate proportion of practically absent genes, the remaining genes were very lowly expressed and their summary parameters exhibited none of the structures observed in the true data. The \textit{muscat} sample contained no genes made up of only zeroes and ones, and the range of $m$ values was much smaller for sparse genes. It did however contain highly-expressed genes and exhibited similar relationships between summary parameters. 

From these observations, we can outline particular validation tasks for which these simulated samples are inappropriate. When assessing a quality control method's ability to filter out very lowly expressed genes, the \textit{muscat} sample's lack of any such genes would make it unsuitable. Similarly, the low gene expression in the \textit{Hierarchicell} sample makes it unsuitable for validating any methods which make use of the information associated with highly-expressed genes. Neither sample captured the relationship between sparsity, mean and kurtosis. Consequently, they could not inform us as to whether the kurtosis in real data compromises the effectiveness of any inferential method.

It is important to note that critiquing a simulated dataset is not critiquing the algorithm that generated it. While our visualisations highlight differences between a dataset and simulations intended to mimic it, this does not discount the possibility that a better simulation could be generated by changing the simulation parameters. Additionally, while we can highlight disparities between simulated and true data, apparent concurrence between them is not proof of similarity. Other descriptive approaches may highlight new dissimilarities, and all visualisations inevitably contain less information than the samples they describe.

When using analyses that lack robust validation or rigorously proven mathematical properties, we must rely on a plurality of visualisations with well understood properties. Our method extends the suite of available techniques, providing a robust system of visualisations built on meaningfully defined summary statistics, without requiring the user to experiment with subjective tuning parameters. This means that when visualisations highlight differences between two datasets, they can be characterised simply and objectively.

\subsection{\textit{T.\ brucei} data}

The detection of the outlier gene in the \textit{T.\ brucei} data demonstrated our method being insightful when a few genes violate our model assumptions. Due to the suitability of our model for the majority of genes, the genes to which our model was not applicable are worth investigating as outliers. We have therefore established  that Wasserstein-based diagnostics can find genes with patterns of expression that differ substantially from other genes. 

\subsection{Potential applications}
We have shown that ZINBGT visualisations can be added to the suite of exploratory methods. Unlike UMAP and t-SNE, its summary statistics have straightforward interpretations in terms of a simple model, allowing for direct, objective comparisons of datasets. This is accomplished without choosing parameter settings, avoiding the subjectivity of tuning parameters. What follows is an outline of some of the applications for our method, some of which have been demonstrated in this paper.

The analysis of the \textit{T. brucei} dataset demonstrated that Wasserstein diagnostics are an efficient means of detecting outlier genes; our method also allows for outlier genes to be detected through unusual parameter values, visible by their placement on figures well-separated from other genes. In the CD14 data, we showed that fitted parameter values allowed the detection of genes that were not meaningfully expressed in a given sample.
This ties into ZINBGT's usefulness when comparing two datasets (as seen comparing the CD14 dataset against two simulations), since the visualisations could be used to assess the impact of quality control and other transformations upon a dataset. While we have not compared datasets with shared genes in this paper, we have developed visualisations that allow for a sophisticated contrast of gene expression across samples by showing how each gene's parameter estimates differ across samples.

Our visualisations can be used to assess the verisimilitude of simulated datasets, allowing users to assess the validity of any benchmarking that uses them as a ground truth. More broadly, our method improves the understanding of the distribution of RNA across cells; this is relevant to the issue of justifying model assumptions. We have seen that zero-inflated negative-binomial distributions can be highly applicable for the majority of genes in a dataset, but the frequent inclusion of a geometric component suggests that either a more flexible model needs to be adopted, or that a dataset needs to be transformed.

Regarding Exploratory Data Analysis more broadly, we have shown that designing mixture distributions to resemble particular empirical distributions provides parameter estimates that act as meaningful summary statistics. Combining this with penalising complexity allows for the use of novel 2D histograms which distinguish between boundary and non-boundary values, defined on a fine grid that maintains the high resolution of scatter plots with the ability to ascertain point density seen in heatmaps. We also demonstrated that the Wasserstein distance and the bootstrapped $p_B$-values quantify similarity between empirical distributions and their fitted models, which helps assess model fit and detect outliers.
Extensions of this work could either broaden it within the realm of exploratory statistics, in which ease for the user is prioritised, or develop it into an inferential method. In either case, it would be advantageous to account for the relationships between genes and consider the variability between cells.

\section{Competing interests}
No competing interest is declared.

\section{Author contributions statement}
Toby Kettlewell (Conceptualisation [lead], Methodology [lead], Software [lead], Writing -- original draft [lead] \& Writing -- review and editing [lead]), Yiyi Cheng (Data curation [lead], Resources [lead] \& Writing -- original draft [support]), Thomas D. Otto (Conceptualisation [support], Supervision [support] \& Writing -- review and editing [support]), Vincent Macaulay (Conceptualisation [support], Methodology [support], Supervision [equal] \& Writing - review and editing [support]) and Mayetri Gupta (Conceptualisation [support], Methodology [support], Supervision [equal] \& Writing -- review and editing [support]).

\section{Acknowledgments}
This work was supported by the Engineering and Physical Sciences Research Council [EP/T517896/1] and the China Scholarship Council (Grant No.\ 202208060071). We would like to thank the members of the Thomas Otto lab for their feedback on the project throughout its development, particularly Domenico Somma for his recommendations during the conceptualisation phase.

\bibliography{reference}
\bibliographystyle{my_apalike}

\clearpage

\appendix

\setcounter{page}{1}
\setcounter{figure}{0}
\setcounter{table}{0}
\setcounter{equation}{0}

\floatname{algorithm}{Supplementary Algorithm}
\renewcommand{\tablename}{Supplementary Table}
\renewcommand{\figurename}{Supplementary Figure}
\renewcommand\thefigure{\thesection.\arabic{figure}} 
\renewcommand\thetable{\thesection.\arabic{table}} 
\renewcommand{\theequation}{\thesection.\arabic{equation}}

\section{Parameter estimation}
\label{app:param_est}

We aim to estimate the parameters, $\boldsymbol \theta = (p_0, p_1, p_2, m, d, \mu_g)$, for ZINBGT using maximum-likelihood estimation when we have a vector of gene-level RNA data, $\mathbf x = (x_1, \dots, x_n)^{T}$. 
The likelihood for mixture models does not permit closed-form expressions for the maximum-likelihood estimators (m.l.e.s) of the model parameters, and direct numerical maximisation of the likelihood function is frequently challenging. In order to maximise the likelihood, $L(\boldsymbol{\theta};\mathbf{x})$, we use the Expectation Maximisation algorithm \citep{dempster_maximum_1977} and introduce a random vector of component-membership terms, $\mathbf{z} \in \{0,1,2\}^n$, where $z_i = k$ if cell $i$ belongs to component $k \in \{0,1,2\}$. It is worth noting that we are not suggesting that any datasets are generated by three separate biological processes. Rather, we are proposing that the empirical mass functions of mRNA counts over cells resemble weighted averages of three standard distributions. These components are a constant zero, a hurdle negative-binomial, and a hurdle geometric distribution, with the following mass functions:
\begin{align}
    f_0(x) &= \delta_0(x), \label{eqn:f0_def}\\
    f_1(x;m, d) &= 
    \left \{ 
    \begin{matrix}
        \frac{1-\delta_0(x)}{d^{\frac{m}{d-1}}-1}
	\frac{\Gamma(x+\frac{m}{d-1})}{x!\Gamma(\frac{m}{d-1})} 
	\frac{(d-1)^x}{d^x} & m>0, d>1 \label{eqn:f1_def} \\ 
        \frac{1-\delta_0(x)}{1-e^{-m}}\frac{e^{-m}m^x}{x!} & m>0, d=1 \\ 
        \delta_1(x) & m=0
    \end{matrix} \right. \quad,\\
    f_2(x;\pi_g) &=  \big(1-\delta_0(x)\big)\frac{ \mu_g^{x-1}}{(1+\mu_g)^{x}}. \label{eqn:f2_def}
\end{align}
where $\delta_i(x)$ equals 1 when $x=i$, and 0 otherwise, i.e., $\delta_i(x) = \mathbb{I}[x=i].$ 
Note that when using the conventional parametrisation of the negative-binomial distribution in terms of $r$ and $p$, it is common for the density function to be defined in terms of factorial functions rather than gamma functions. Defining the density in terms of gamma functions allows for $r$ to be a non-integer. As we have replaced $r$ with $\tfrac{m}{d-1}$, which will not generally be an integer, we have also adopted this convention.

If we treat the component membership of the $i$-th cell as observed then the complete data likelihood becomes
\begin{equation}
   L(\boldsymbol{\theta};\mathbf{x},\mathbf{z}) = 
    \prod_{i=1}^n \prod_{k=0}^2 f_k(x_i ; \boldsymbol{\theta})^{\mathbb{I}[z_i=k]}.
\end{equation}
The log of the complete data likelihood is equivalent with respect to optimisation:
\begin{equation}
   \log \big( L(\boldsymbol{\theta};\mathbf{x},\mathbf{z}) \big) = 
   \sum_{i=1}^n \sum_{k=0}^2 {\mathbb{I}[z_i=k]} \log \big (f_k(x_i ; \boldsymbol{\theta})\big ).
\end{equation}

If we knew $\mathbf{z}$, this would collapse into three separate m.l.e.\ problems, one for each component, as elements of $\boldsymbol \theta$ are not shared between components. Similarly, if we knew $\boldsymbol \theta$, we could estimate $\mathbf z$ using Bayes' theorem. The conceit of EM is that by choosing an initial approximation of either $\boldsymbol \theta$ or $\mathbf z$, we can iteratively estimate each from the other, repeating until no meaningful change in parameter estimates is seen. One iteration of this algorithm consists of calculating:
\begin{equation}
\begin{split}
\label{eqn:EM_theta_update}
\boldsymbol \theta ^{(t+1)} & = 
\underset{\boldsymbol \theta}{\arg \max}
    \Big( \mathbb{E}_{\mathbf z|\mathbf{x}, \boldsymbol{\theta}^{(t)} }
   \big[ \log \big(L
   	( \boldsymbol{\theta} ;\mathbf{x} ,\mathbf z)
   \big)
   \big]
   \Big) \\
    & =
   \underset{\boldsymbol \theta}{\arg \max} 
   \left( 
   \sum_{i=1}^n \sum_{k=0}^2
   \mathbb{P}[z_i=k| x_i, \boldsymbol{\theta}^{(t)}] \log \big (f_k(x_i; \boldsymbol{\theta})\big )
      \right),
\end{split}
\end{equation}
for $t=0, 1, \dots$, where $\boldsymbol \theta ^{(t+1)}$ is the estimate of $\boldsymbol \theta$ at iteration $t+1$. This is accomplished by first determining $\gamma_{ik}^{(t+1)}:=\mathbb{P}[z_i=k| x_i, \boldsymbol{\theta}^{(t)}]$ using Bayes' theorem. This is the Expectation step. Then, in the Maximisation step, the $\gamma_{ik}^{(t+1)}$ terms are entered into (\ref{eqn:EM_theta_update}) and $\boldsymbol \theta^{(t+1)}$ found. The following sections outline the details of how these steps are accomplished for ZINBGT, and strategies for selecting initial values.

\subsection{Expectation and Maximisation steps}

For all $i \in \{1,\dots,n\}, k \in \{0,1,2\}$, Bayes' theorem gives 
\begin{equation}
\label{eqn:EM_comp_memb_prob}
\begin{split}
\gamma_{ik}^{(t+1)} = 
    \mathbb{P}[z_i = k|x_{i}, \boldsymbol{\theta}^{(t)}] &= 
    \frac{f_k(x_{i}; \boldsymbol{\theta}^{(t)})  \mathbb{P}[z_i = k|\boldsymbol{\theta}^{(t)}] }
         {\sum_{k'=0}^2 f_{k'}(x_{i}; \boldsymbol{\theta}^{(t)})  \mathbb{P}[z_{i}=k'|\boldsymbol{\theta}^{(t)}] }\\ 
    		& = 
    \frac{f_k(x_{i} ; \boldsymbol{\theta}^{(t)}) p_k^{(t)}}
    		{\sum_{k'=0}^2 f_{k'}(x_{i} ; \boldsymbol{\theta}^{(t)}) p_{k'}^{(t)}}
.
\end{split}
\end{equation}

The maximisation in (\ref{eqn:EM_theta_update}) must be performed while imposing the restriction $\sum_{i=0}^2 p_i^{(t+1)} = 1$, which can be achieved using the Lagrange multiplier method. The Lagrangian function, making use of (\ref{eqn:EM_comp_memb_prob}), is
\begin{multline}
 \label{eqn:zinbgt_lagrange}
 \mathcal{L} = 
    \lambda (p_0 + p_1 + p_2 - 1) + 
 	\sum_{i=1}^n
 	\Bigg \{ 
 		\gamma_{i0}  \log (p_0) \\  +
 		\gamma_{i1} 
 		\Bigg[	
 			\log(p_1) +
 			\log \Bigg ( \Gamma \bigg (x_i + \frac{m}{d-1} \bigg) \Bigg) 
 			- \log (x_i!) 
 			- \log \Bigg ( \Gamma \bigg(\frac{m}{d-1} \bigg) \Bigg )
 			+ x_i \log(d-1) 
 			- x_i \log(d) 
 			- \log \left( d^{\frac{m}{d-1}}-1 \right)
 		\Bigg] \\  +
 		\gamma_{i2} \big[\log(p_2) + (x_i-1) \log(\mu_g) - x_i \log(\mu_g+1) \big] 
 	\Bigg \}.
 \end{multline}
 \noindent 
Note that here we have dropped the Kronecker delta terms. If we assume that we will only allocate the zero values to the zero component and all positive values to the non-zero components, then the Kronecker terms will invariably equal one and they can be omitted without consequence.

By differentiating (\ref{eqn:zinbgt_lagrange}) with respect to the desired parameters, including $\lambda$, we are able to find maximum-likelihood estimators using $\Gamma^{(t)}= \{\gamma^{(t)}_{ik}\}_{i=1,k=0}^{n,2}$. The parameters $m$ and $d$ must be estimated together numerically, as hurdle negative-binomial distributions do not have closed form m.l.e.s. This is achieved using R's numerical optimisation function, \texttt{optim}, to maximise the terms in (\ref{eqn:zinbgt_lagrange}) containing $m$ and $d$.
Differentiating and setting the other parameters to their maximum-likelihood estimates yields:
\begin{align}
    \label{eqn:EM_prob_mle}
        \left. \frac{\partial \mathcal{L}}{\partial p_k }\right|_{\boldsymbol \theta =  \hat{\boldsymbol \theta },\  \lambda= \hat \lambda} = 0 =
        {}& 
        \sum_{i=1}^n \frac{\gamma_{ik}}{\hat p_k} + \hat \lambda,\; k \in \{0,1,2\},
        \\
    \label{eqn:EM_lambda_mle}
        \left . \frac{\partial \mathcal{L}}{\partial \lambda } \right|_{\boldsymbol \theta =  \hat{\boldsymbol \theta },\ \lambda= \hat \lambda} = 0 =
        {}& 
        \hat p_0 + \hat p_1 + \hat p_2 - 1,
        \\
    \label{eqn:EM_p_geom_mle}
        \left . \frac{\partial \mathcal{L}}{\partial \mu_g } \right|_{\boldsymbol \theta =  \hat{\boldsymbol \theta },\ \lambda= \hat \lambda} = 0 =
        {}& 
        \sum_{i=1}^n \gamma_{i2} \left( \frac{x_i-1}{\hat \mu_g} - \frac{x_i}{1+\hat \mu_g} \right).
\end{align}

\noindent Rearranging (\ref{eqn:EM_prob_mle}) gives $\hat \lambda \hat p_k = -\sum_{i=1}^n \gamma_{ik}, \forall k$. By summing these equations over $k$ and using (\ref{eqn:EM_lambda_mle}), $\hat \lambda = -n$ and therefore $\hat p_k = \frac 1 n \sum_{i=1}^n \gamma_{ik}.$
Rearranging (\ref{eqn:EM_p_geom_mle}) returns 
$\hat \mu_g= \frac{\sum_{i=1}^n \gamma_{i2} (x_i-1)}{\sum_{i=1}^n \gamma_{i2}}$, which is the weighted mean of the count values, weighting each count by the probability it came from the geometric component and adjusting for the use of a hurdle distribution by decreasing the count values by one.

\subsection{The Poisson case}

While the Poisson distribution has the mean of the observations as the m.l.e.\ of its parameter, the m.l.e.\ for the parameter of a hurdle Poisson does not have a closed form. As a result, the ZINBGT model's m.l.e.\ for $m$ cannot be written in closed form in the Poisson case ($d=1$). When modifying (\ref{eqn:EM_theta_update}) to reflect the negative-binomial component having a Poisson distribution, the Lagrangian function is 
  \begin{multline}
 \mathcal{L} = \lambda (p_0 + p_1 + p_2 - 1)\; \\ + 
 	\sum_{i=1}^n
 	\Big\{ 
 		\gamma_{i0}  \log (p_0)  + 
 		\gamma_{i1} 
 		[	
 			\log(p_1) - m +x_i \log(m) - \log(x_i!) - \log(1 - e^{-m})
 		] \\ + 
 		\gamma_{i2} [\log(p_2) + \log(\pi_g) + (x_i-1) \log(1-\pi_g)]
 	\Big \},
 \end{multline}
and so, at the m.l.e.,
 \[
	\left . \frac{\partial \mathcal{L}}{\partial m } \right|_{\boldsymbol \theta =  \hat{\boldsymbol \theta }}= 
	\sum_{i=1}^n \gamma_{i1}\left[-1 + \frac{x_i}{\hat m} -\frac{e^{-\hat m}}{1-e^{-\hat m}}\right]  = 0.
 \]
Introducing $\tilde n = \sum_{i=1}^n \gamma_{i1}$, $\tilde x = \sum_{i=1}^n \gamma_{i1} x_i / \tilde n$ gives:
\begin{equation}
\label{eqn:hurd_m_mle}
    \hat m \frac{1}{1-e^{-\hat m}} = \tilde x.
\end{equation}
As $\hat m$ goes to infinity, $\hat m \frac{1}{1-e^{-\hat m}}$ approaches $\hat m$. This reflects the m.l.e.\ of the hurdle Poisson parameter approaching the m.l.e.\ of the non-hurdle version (the weighted mean) as the probability of the random variable being zero approaches zero.

While no explicit solution of (\ref{eqn:hurd_m_mle}) exists (Bronstein \textit{et al.}, 2008), we can approximate the value for large $\tilde x$ by taking $\hat m = \tilde x$. For smaller values, a grid of $m$ values can be selected, and their associated $\tilde x$ values calculated. Following this, linear interpolation allows for estimation of $\hat m$ from observed small $\tilde x$ values. A trade-off between speed and accuracy was made by choosing a cut-off of $\tilde x = 30$, with 3000 grid values placed logarithmically between a minimum parameter value and 30. The minimum parameter value was chosen to prevent numerical issues found with the negative-binomial distribution mass function in R when parameter values are small. The associated $\tilde x$ value for this minimum was one, as this is the smallest possible value that can be observed.

\clearpage
\section{Initialisation strategy experiments}
\label{app:init_strat}

\begin{table}[h!]
\begin{center}
\begin{tabular}{ r r r r r r r r}
\hline
Even & Exponential & Median & Random\\ 
 \hline
19 & 32 & 63 & 27 \\
\hline
\end{tabular}
\end{center}
		\caption[Table of the number of genes for which each initialisation was optimal in terms of maximising likelihood.]{Table of the number of genes for which each initialisation was optimal in terms of maximising likelihood.}
\label{table:init_lik_best_1_rand}
\end{table}

\begin{figure*}[!h]%
\centering
{\includegraphics{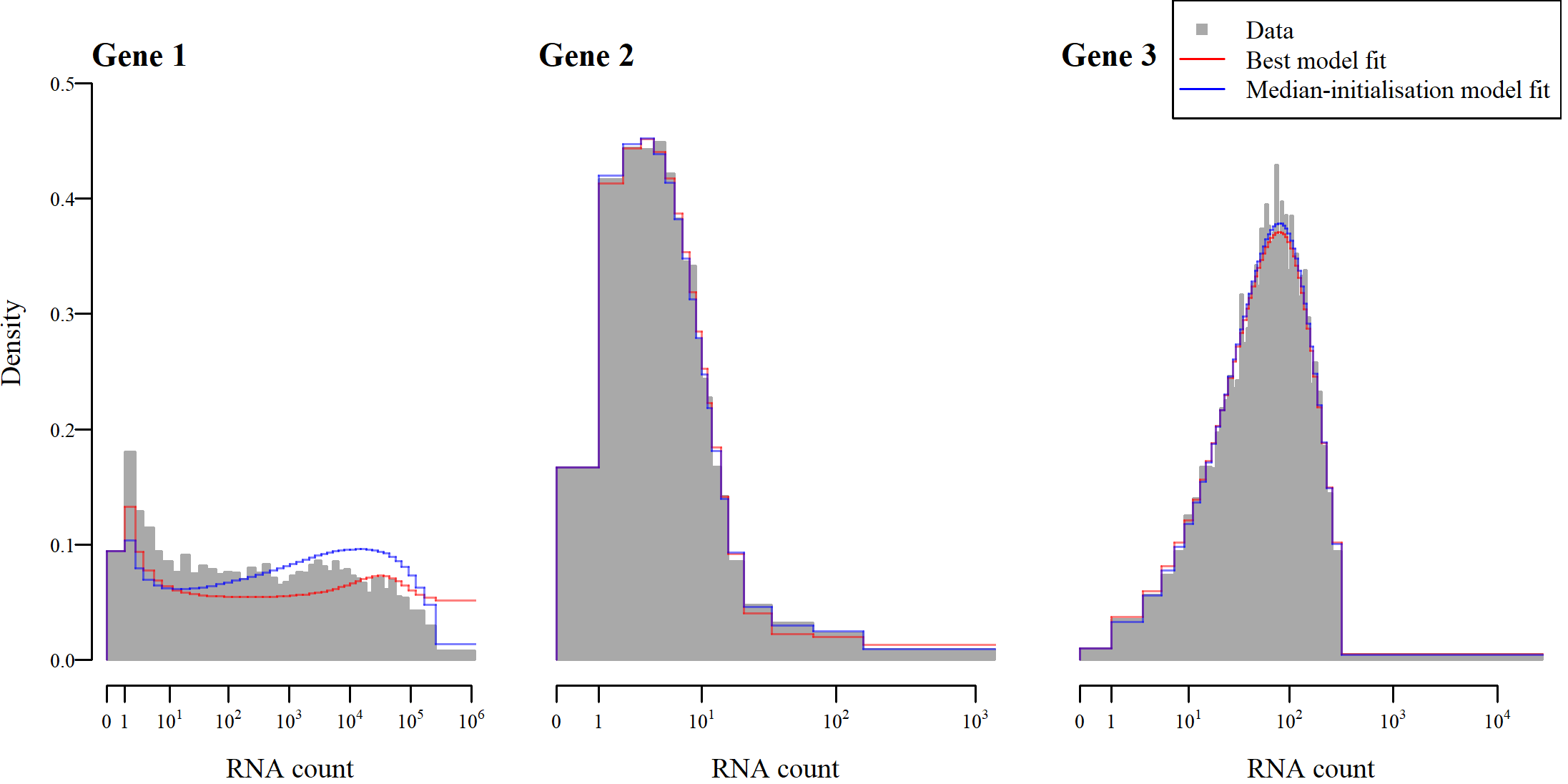}}
\caption{Histograms of the count values of the three simulated genes for which median initialisation was not optimal, and for which the difference in likelihood between the median initialisation and best strategy was the greatest. The lines on each histogram represent that gene's model distribution for the median initialisation and best strategy.}
\label{fig:init_strat}
\end{figure*}

A number of initialisation strategies were devised and tested. Model fitting for each gene's zero component is trivial, as the m.l.e.\ of $p_0$ is the proportion of that gene's RNA counts that are zero. We therefore initialise $p_0$ to its m.l.e.\ and it remains unchanged throughout model fitting. The differences between strategies relate only to the initial conditions of parameters concerned with non-zero values or the component-membership probabilities of non-zero count values. 

The \textit{even} strategy assigns all non-zero counts an equal probability of being from the negative-binomial or geometric components. In designing our mixture distribution, we intend for the geometric component to account for the larger counts in the tail, allowing for greater flexibility regarding kurtosis than is possible with zero-inflated negative-binomial distributions. To achieve this, we can initialise allocations such that smaller values are more likely to belong to the negative-binomial component, while the larger values are associated more strongly with the geometric. 
We have designed the \textit{exponential} allocation to reflect this. 
Here, we initialise the probability of a non-zero count value $x$ belonging to the negative-binomial component as $10^{x-1}$. Consequently, its probability of belonging to the geometric component is $1-10^{x-1}$.
The \textit{median} allocation tries to force the negative-binomial component to take lower values than the geometric by initialising $m$ as the median of the non-zero counts, while $\mu_g$ is initialised to the maximum observed count. The dispersion parameter, $d$, is initialised as the ratio of the variance of the observed non-zero counts to the initial value of $m$. If this is less than one, it is instead set to one. The proportion mixture parameters are assigned: $ p_0 = |\{X_i|X_i = 0\}|/n$, $p_1=p_2=(1-p_0)/2$. 
 
These initialisations are all deterministic, i.e., they depend only on the data. In addition, we tested random initialisations, for which the initial allocation of non-zero values is generated under $\mathbb{P}[X_i \sim \text{Hurdle negative-binomial}|X_i=x, X_i \neq 0] \sim \text{U}[0,1]$. We also experimented with running the random initialisation multiple times. We found that this did not lead to us reaching any different conclusions, so it is not discussed here.

As the purpose of the EM algorithm is to maximise the likelihood function, we can compare the relative suitability of each strategy on a particular vector of count data by comparing the likelihoods of their parameters after model fitting. These comparisons will be performed on a simulated dataset made up of 10,000 cells, with the 809 gene parameter values taken from parameter estimates calculated using the CD14 data. In order to ensure that the genes sampled cover a range of plausible values, 200 were taken from the five combinations of simplifications for which model fitting is non trivial. As only nine genes in the CD14 data were fitted with a zero-inflated geometric distribution -- having dropped the negative-binomial entirely -- they are all used and a full set of 200 is not possible.

Having applied model fitting to each of the 809 genes using the four different initialisations, 141 genes were found with likelihoods differing according to initialisation strategy. Supplementary Table \ref{table:init_lik_best_1_rand} shows how many of the genes had the highest likelihood for each deterministic strategy and one of the random initialisations. While no strategy appears to be categorically better than the others, the \textit{median} strategy outperformed the others substantially.

This alone is not sufficient for us to state that the \textit{median} strategy performed the best on this sample, as it does not account for how different the likelihoods are, only which strategy had the largest. In those circumstances where another strategy beats the \textit{median} strategy, it is important to establish whether the fitted parameter values are still able to faithfully describe the data. Supplementary Figure \ref{fig:init_strat} visualises the count data of the three genes with the greatest disparity in likelihood between the \textit{median} initialisation and strategy that was best. The lines overlaid compare the model distributions following the \textit{median} and the best strategy. 
We do see a substantial difference between initialisations for Gene 1, but we also see that neither strategy has been able to produce parameters that faithfully describe the data. Therefore, while the \textit{median} strategy was not the best in terms of likelihood, no strategy resulted in appropriate parameter values. For the other two genes, the differences in model distributions are minor, suggesting that for these genes -- and likely the others -- the suboptimal \textit{median} parameter estimates are comparable to the best for the purpose of summarisation. Based on this experiment, we have decided to use the \textit{median} initialisation.

\clearpage
\section{Wasserstein diagnostics experiments}
\label{app:wass_experiments}

\subsection{Performance of diagnostics when the model assumptions hold}

\begin{figure*}[!h]%
\centering
{\includegraphics{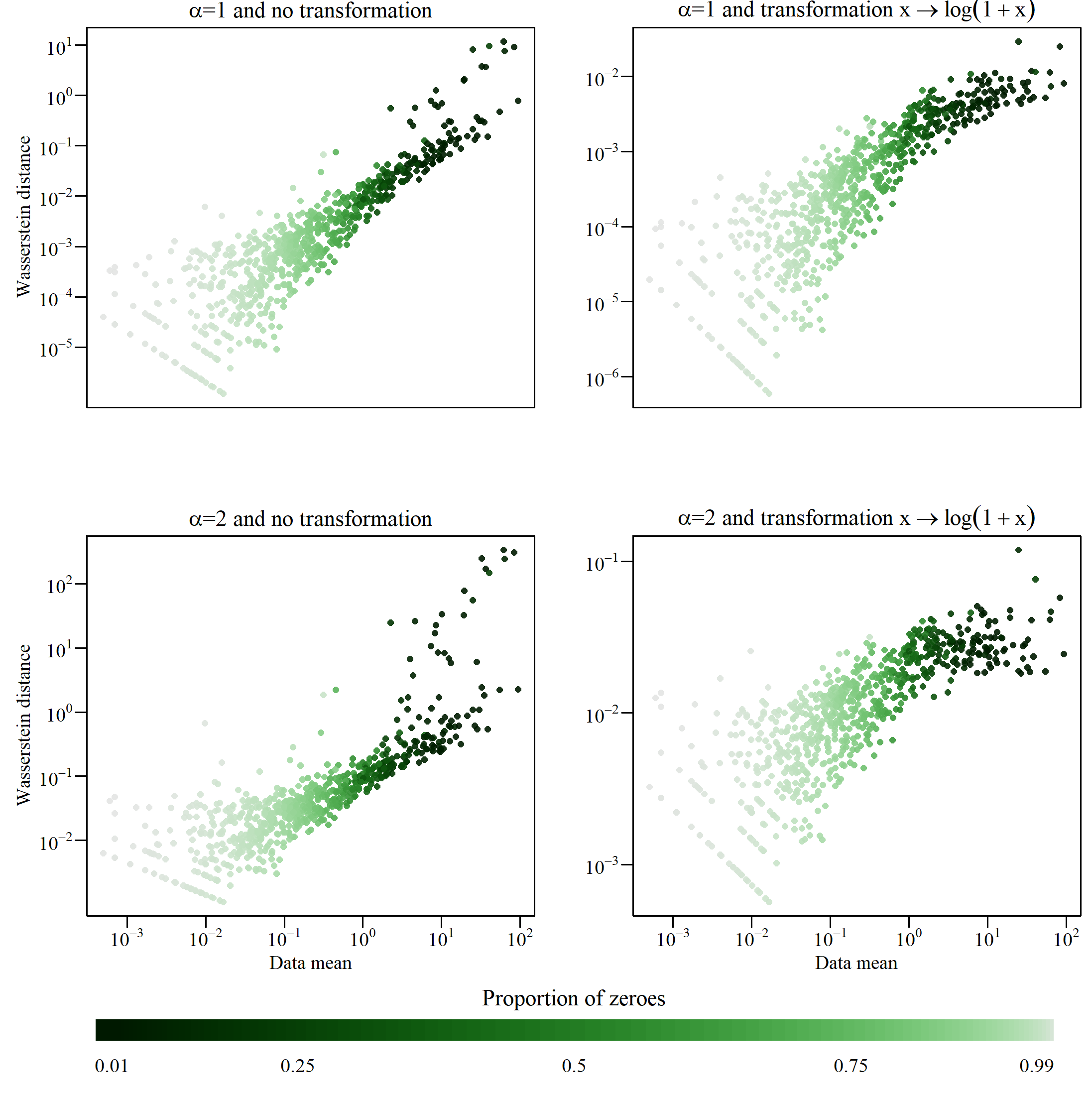}}
\caption{Scatter plots for comparing the sample mean and Wasserstein distance for each gene in the model-validation sample, with the colour reflecting the proportion of zeroes. For each scatter plot, the Wasserstein distances are either calculated with the cost parameter $\alpha$ equal to one or two, and with either the original count values or the count values transformed by $\texttt{log1p}$.}
\label{fig:wass_mean_dist}
\end{figure*}

\begin{figure*}[!h]%
\centering
{\includegraphics{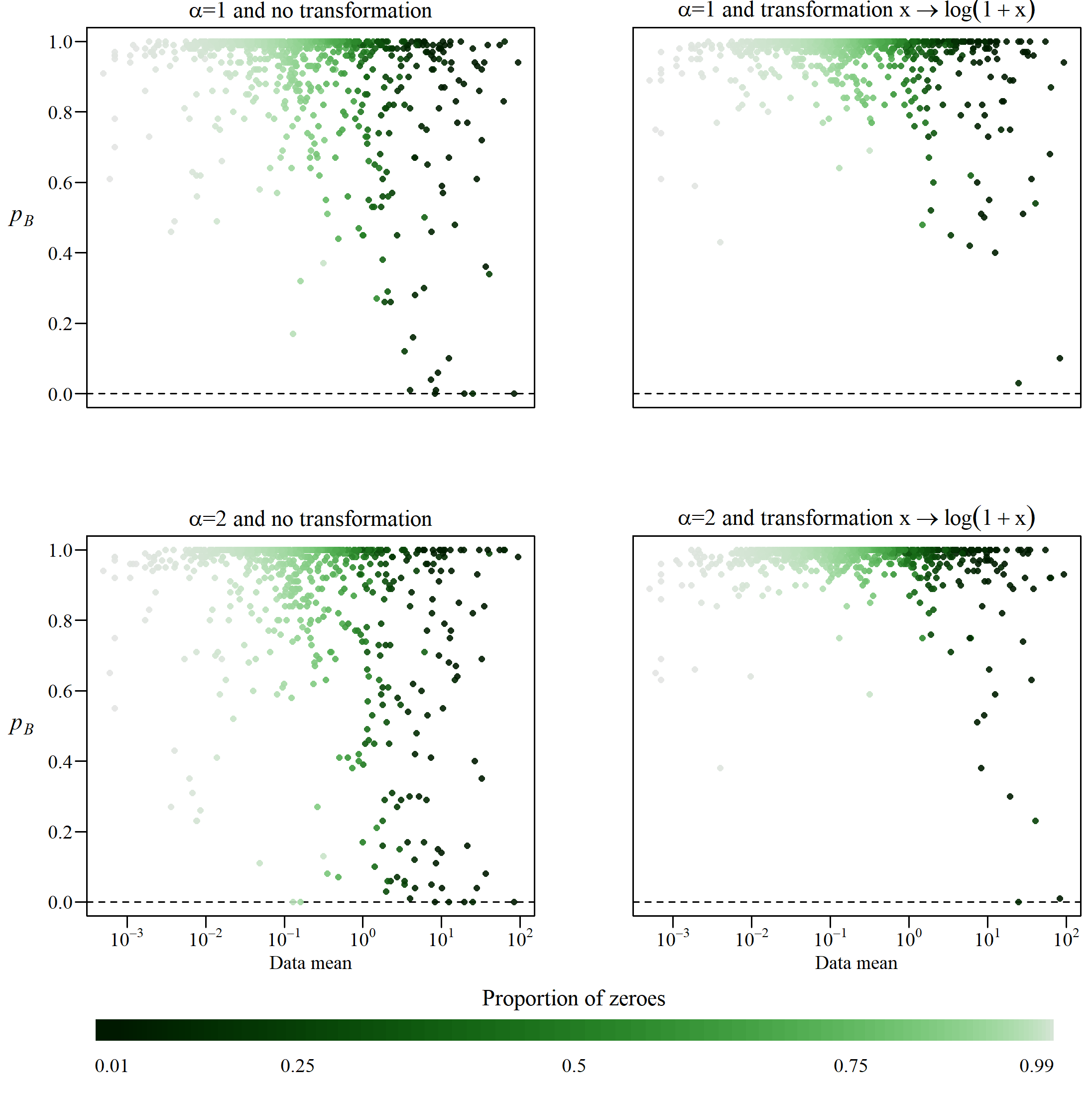}}
\caption{Scatter plots for comparing the sample mean and $p_B$-value for each gene in the model-validation sample, with the colour reflecting the proportion of zeroes. For each scatter plot, the $p_B$-values are either calculated with the cost parameter $\alpha$ equal to one or two, and after either not transforming the count values or applying the $\texttt{log1p}$ transformation.}
\label{fig:wass_mean_pb}
\end{figure*}

Comparisons of the four Wasserstein diagnostics strategies were first performed on the simulated data discussed in Appendix \ref{app:init_strat}. We explored $\alpha = 1$ and $\alpha = 2$ as the former, named the \textit{earth mover's distance}, is popular, and mathematical proofs usually concern the latter (e.g, Panaretos and Zemel, 2019). Supplementary Figures \ref{fig:wass_mean_dist} and \ref{fig:wass_mean_pb} show the relationship between mean count value, Wasserstein distance and $p_B$-value. In Supplementary Figure \ref{fig:wass_mean_dist} we see that the sample mean and Wasserstein distance are positively correlated regardless of strategy, but that this correlation becomes weaker for larger means when the \texttt{log1p} transformation is used. We also notice that the \texttt{log1p} transformation of the count values resulted in fewer outlier genes with larger than average Wasserstein distances for their mean count value. As all of the data in this sample was generated by ZINBGT distributions, the presence of genes that look like outliers is misleading. If a sample were to also include genes for which ZINBGT was suitable, the presence of well-fitted genes that resemble outliers would complicate the exploration of the problematic genes.

Supplementary Figure \ref{fig:wass_mean_pb} shows a weaker relationship between sample means and $p_B$-values, particularly when using a \texttt{log1p} transformation. This motivates the use of $p_B$-values rather than Wasserstein distance to detect outlier genes. Comparing the four strategies, \texttt{log1p} transformation was better at not falsely flagging genes as problematic, with a cost parameter of one resulting in no genes flagged with a $p_B$-value of zero, and a parameter of two flagging one with another nearby.

\begin{figure}[!h]%
\centering
{\includegraphics{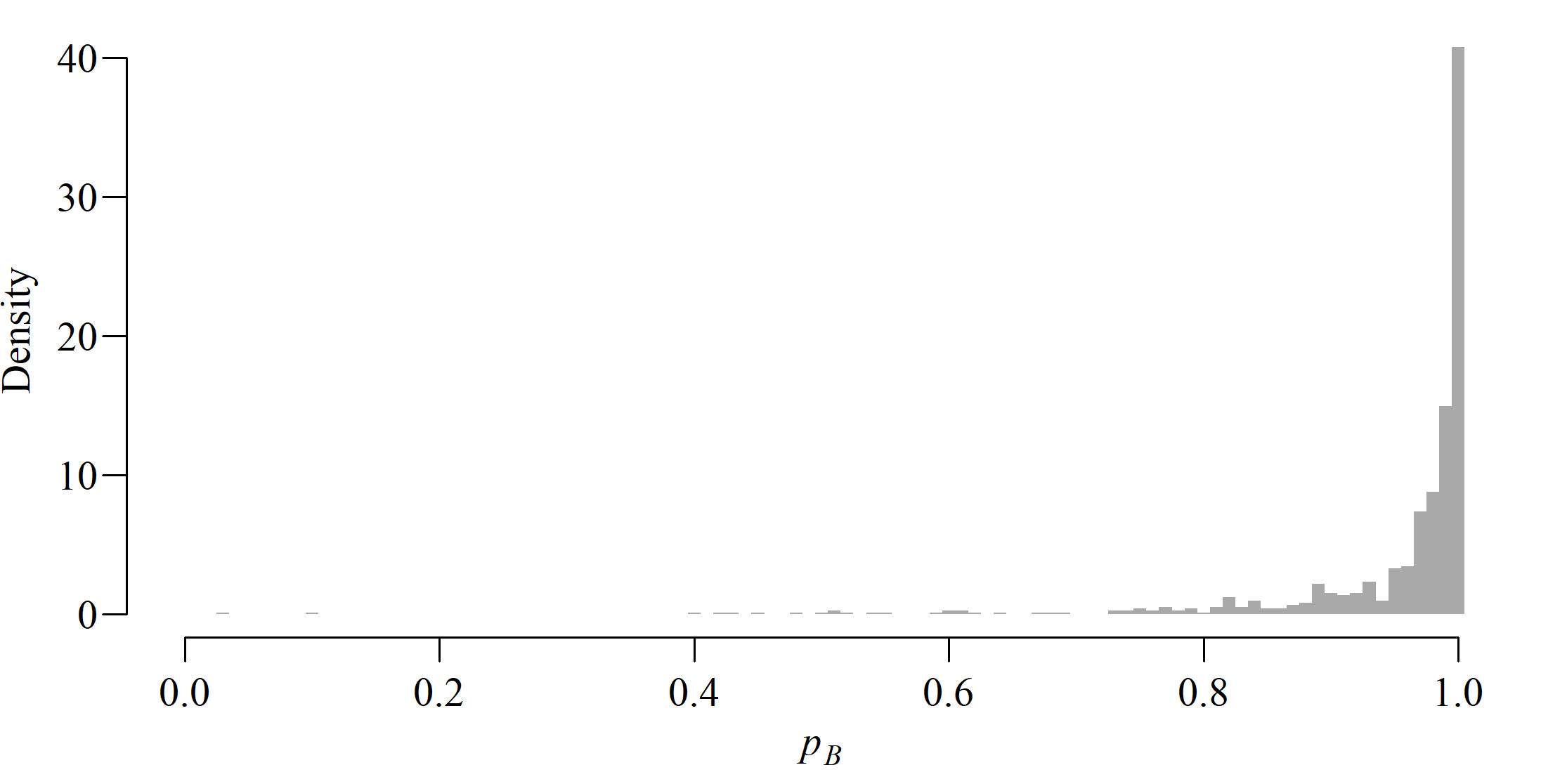}}
\caption{Histogram of the $p_B$-values calculated on the ZINBGT-generated validation dataset with a cost parameter, $\alpha$, of one and applying a $\texttt{log1p}$ transformation.}
\label{fig:wass_1_log1p_pb}
\end{figure}

\begin{table}[t]
    \centering
    \begin{tabular}{crrrr}
         &  \multicolumn{4}{c}{Transformation}\\
         \cmidrule(lr){2-5}
         &  \multicolumn{2}{c}{None}&  \multicolumn{2}{c}{\texttt{log1p}}\\
         \cmidrule(lr){2-3} \cmidrule(lr){4-5}
         $p_B$-value&  $\alpha=1$&  $\alpha=2$&  $\alpha=1$& $\alpha=2
$\\
         \cmidrule(lr){1-1} \cmidrule(lr){2-2} \cmidrule(lr){3-3} \cmidrule(lr){4-4} \cmidrule(lr){5-5}
         $\{0\}$&  435&  427&  467& 420\\
         $(0,1)$&  513&  467&  401& 380\\
         $\{1\}$&   52&  106&  132& 200\\
    \end{tabular}
    \caption[Table of the number of simulated, non-ZINBGT genes assigned a particular significance.]{Table of the number of simulated, non-ZINBGT genes assigned a particular significance. Each column represents a diagnostic strategy.}
    \label{tab:misspec_signif}
\end{table}

\subsection{Performance of diagnostics under violation of model assumptions}

While these experiments suggest that the use of the \texttt{log1p} transformation is better for avoiding falsely highlighting genes -- with perhaps weak evidence favouring $\alpha=1$ -- there is still the issue of how these strategies compare at correctly identifying outliers. 

To explore this, we created a simulated dataset of 1000 genes and 10,000 cells that was not generated from ZINBGT distributions. As the family of distributions that violate our model assumptions is infinite, we focussed on a family of distributions that reflected a particular pattern of expression we might anticipate in a scRNA-seq dataset: mixtures of two non-hurdle negative-binomial distributions. This reflects the popularity of negative-binomials when modelling scRNA-seq data, while allowing for a bimodality that ZINBGT cannot capture. For each gene $j$ and cell $i$, the simulated count value $X_{ij}$ is independently generated such that

\begin{equation*}
\mathbb{P}[X_{ij}=x|\rho_j, m_{j1}, m_{j2}, d_{j1}, d_{j2}] =
    \rho_j \text{NB}(x; m_{j1}, d_{j1}) + (1-\rho_j) \text{NB}(x; m_{j2}, d_{j2}),
\end{equation*}
where the parameters for each $j$ are independently generated by:
 $$m_{j1} \sim \text{Exp}\left(\frac{1}{10}\right); 
    m_{j2}|m_{j1} \sim m_{j1}\Bigg(1+\text{Exp}\bigg(\frac{1}{10}\bigg)\Bigg); 
    d_{j1}, d_{j2} \sim 1+\text{Exp}\left(\frac{1}{10} \right);
    \rho_j \sim \text{Beta}(2,2).$$ 

After model fitting was applied to this simulated dataset, the $p_B$-values for each gene under each strategy were calculated, the values of which are shown in Supplementary Table \ref{tab:misspec_signif}. It shows that the \texttt{log1p} with $\alpha=1$ assigned more of the genes to $p_B$-values of zero, which is consistent with being better than the other strategies identifying assumption-violating genes as such.

A visualisation of the genes that were assigned higher $p_B$-values suggested that their not being highlighted as problematic was not concerning. Note that the ZINBGT family of distributions and the family of mixtures of two negative-binomial distributions overlap -- in part because geometric distributions are special cases of negative-binomials -- which results in samples being generated that are consistent with, and describable as, ZINBGT generated. These higher $p_B$-value genes were found to be well described by ZINBGT distributions, with the remaining genes being unambiguously bimodal.

Due to these results, the \texttt{log1p} with $\alpha=1$ strategy was adopted as the preferred strategy when calculating the Wasserstein distance and $p_B$-values.

\clearpage

\section{Pseudocode for calculation of $p_B$-values}

\label{appsub:pb_pseudo}

\renewcommand{\algorithmicrequire}{\textbf{Input:}}
\renewcommand{\algorithmicensure}{\textbf{Output:}}

\begin{algorithm}[H]
\caption{Pseudocode for calculating a $p_B$-value.}
\label{alg:wass_boot}
\begin{algorithmic}[]
\Require $W_D$: Wasserstein distance between real data and fitted model
\renewcommand{\algorithmicrequire}{\phantom{\textbf{Input:}}}
\Require $\boldsymbol{\hat{\theta}}$: Parameter values calculated on real data
\Require $n$: Sample size of real data
\Require $k$: Number of simulated samples evaluated  (100 by default)
\Ensure $p_B$
\State Initialise counter, $c \leftarrow 0$, for the number of simulated samples with Wasserstein distances larger than $W_D$ 
\For{$i = 1, \dots, k$}
    \State $\textbf{X} \leftarrow n$ values generated from ZINBGT($\boldsymbol{\hat{\theta}}$)
    \State $W_B \leftarrow$ Wasserstein distance between $\textbf{X}$ and ZINBGT($\boldsymbol{\hat{\theta}}$)
    \If{$W_B \ge W_D$}
        \State $c \leftarrow c+1$
    \EndIf
\EndFor
\State $p_B \leftarrow c/k$
\State \textbf{return} $p_B$ 
\end{algorithmic}
\end{algorithm}

\clearpage
\section{Differences between $p_B$-values and $p$-values}
\label{appsub:pb_isnt_p}

While $p_B$-values are analogous to $p$-values, we must emphasise that they are not the same. Before we illustrate why this is the case, it is helpful to provide a conventional definition of a $p$-value. We start with a vector of random observations, $\mathbf{x}$, which we treat as a realisation of some unknown random vector $\mathbf{X}$. We define a null hypothesis,
\begin{equation*}
    \text{H}_0: \mathbf{X} \sim F,
\end{equation*}
where $F$ is a particular, fully-specified probability measure; an alternative hypothesis,
\begin{equation*}
    \text{H}_1: \mathbf{X} \not \sim F;
\end{equation*}
and a test statistic $T$, where $\mathbb{P}[T(\mathbf{X}) \leq T(\mathbf{x})|\text{H}_0]\sim \text{Uni}(0,1)$. The $p$-value is $\mathbb{P}[T(\mathbf{X}) \leq T(\mathbf{x})|\text{H}_0]$, and $H_0$ is rejected if it is small (typically, below 5\%).

The definition of $p_B$-values is superficially similar; we take an observation $\mathbf{x}$, a replicate of $\mathbf{X} \sim F$, where $F$ is unknown. We can define model fitting using a function $\hat F$ which maps from the set of possible samples to distributions within the ZINBGT family. Our null hypothesis is that model fitting was successful, i.e., $\text{H}_0: F \approx \hat F(\mathbf{x})$, and the alternative hypothesis, $\text{H}_1: F \not \approx \hat F(\mathbf{x})$, suggests that either $F$ is not in the ZINBGT family or model fitting was otherwise unsuccessful. Unlike our definition of a $p$-value, this $H_0$ does not state that $\mathbf{x}$ was generated by some fully-specified distribution, rather, that our fitted model, $\hat F(\mathbf{x})$, is \textit{similar} to some unknown generating distribution, $F$. Because ``is similar to" has not been defined and we have not specified $F$, we are not calculating a ``true" $p$-value. 

We will highlight other divergences. For our test statistic, we make use of the Wasserstein distance function, $W$, which maps a pair of probability measures, defined on the same domain, to the non-negative real numbers. Not knowing $F$, we instead compare $\hat F(\mathbf{x})$ to the empirical distribution on $\mathbf x$, $H(\mathbf{x})$, so that our test statistic features $W\Big(\hat F(\mathbf{x}), H(\mathbf{x})\Big)$. We use bootstrapping, because we do not have an expression for the distribution of $W\Big(\hat F(\mathbf{X}), H(\mathbf{X})\Big) \Big|\mathbf{X} \sim F$. Our analogue for $\mathbf{X}$ is a set of bootstrap samples, $\mathbf{z}_i$ for $i=1,\dots, 100$, where $\mathbf{z}_i \sim \hat F (\mathbf{x})$. Our test statistic is 
\begin{equation}
p_B=\frac{1}{100}\sum_{i=1}^{100}
    I\left [W\left (\hat F(\mathbf x), H(\mathbf x) \right) \leq 
    W\left (\hat F(\mathbf x), H(\mathbf z_i) \right) \right]. 
\end{equation}
If it were the case that $F=\hat F(\mathbf{x})$, then $\mathbf{x}$ and all $\mathbf{z}_i$ would be i.i.d., and the test statistic would follow a discrete uniform distribution over $\left \{0, \frac{1}{100}, \dots,\frac{100}{100} \right \}$. Instead, $\mathbf{x} \sim F$ and $\mathbf{z}_i \sim \hat F(\mathbf{x}), \forall i = 1, \dots, 100.$  Because of this, we expect the distributions of $W\left (\hat F(\mathbf x), H(\mathbf x)\right )$ and $ W\left (\hat F(\mathbf x), H(\mathbf z_i)\right)$ to be different, and we are unable to make statements about the distribution of $p_B$. Supplementary Figure \ref{fig:wass_1_log1p_pb} in Appendix \ref{app:wass_experiments} shows $p_B$-values calculated on data generated from a ZINBGT distribution, that is to say, under our null hypothesis. Notably, they are not uniformly distributed over $[0,1]$, exhibiting a mean and mode close to one. Consequently, a gene being assigned a low $p_B$-value is more suggestive of a violation of $H_0$ than a numerically equal $p$-value would be.

\clearpage
\section{CD14 monocyte data simulation}
\label{appsub:sim_alg}

\textit{Hierarchicell} (Zimmerman and Langefeld, 2021)  -- implemented here as in Zimmerman \textit{et al.} (2021) -- and \textit{muscat} (Crowell
\textit{et al.}, 2020) were used to create simulated datasets based on the CD14 monocyte sample. 
Where possible, simulation settings were chosen for each method in order to make the datasets produced more comparable. 
Both datasets consisted of 5000 genes, 20\% of which were differentially expressed over two groups. Each group contained 10 individuals, with a total of 12,000 cells being simulated. Both datasets were generated using the same CD14 sample as the reference data, assigning individuals to one of two groups depending on whether their worst clinical status was labelled as ``Healthy" or ``Moderate". 

\textit{Muscat} provides six ways in which a gene can be differentially expressed across groups, and two ways it can be not differentially expressed. 
For consistency with the \textit{Hierarchicell}-generated data, all differentially-expressed genes were of the type where only the mean expression differs across groups, which \textit{muscat} labels as \texttt{DE}. 
The remaining genes were split evenly between \texttt{EE} (Equivalently Expressed) and \texttt{EP} (Equal Proportion) under \textit{muscat's} typology, i.e., each gene's expression distribution is the same across groups, but genes with \texttt{EE} expression have a unimodal distribution and \texttt{EP} genes have a bimodal distribution. The mean log-fold change was set to 2, the number of clusters to 1 (the cluster being the cells labelled as CD14 in the Covid data), and the \texttt{pair} parameter was set to true. The reference data object was prepared for the data simulation function, \texttt{simData}, using the package's \texttt{prepSim} function with the default parameters.


\textit{Hierarchicell} does not allow for some genes to be differentially expressed and others not. To account for this, the approach described in Murphy and Skene (2022) was adopted: non-differentially expressed genes were simulated with a mean fold change of 1, i.e., no difference across groups, and differentially-expressed genes were generated as usual, here with a mean fold change of 4. Combining the resulting datasets produced a single simulated dataset for which some genes are differentially expressed and others are not. For each of the 20 individuals, 600 cells were generated. The summary of the reference data passed to the simulation function, \texttt{simulate\_hierarchicell}, was created by taking the subset of genes from the reference data that the \textit{Seurat} (Hao \textit{et al.}, 2024) function \texttt{FindVariableFeatures} labelled as variable, and passing it to \textit{Hierarchicell's}   \texttt{filter\_counts\_mod} function, in turn passed to \texttt{compute\_data\_summaries}. The default parameters were used for these functions.

\clearpage
\section{Supplementary Figures}

\begin{figure*}[!h]%
\centering
{\includegraphics{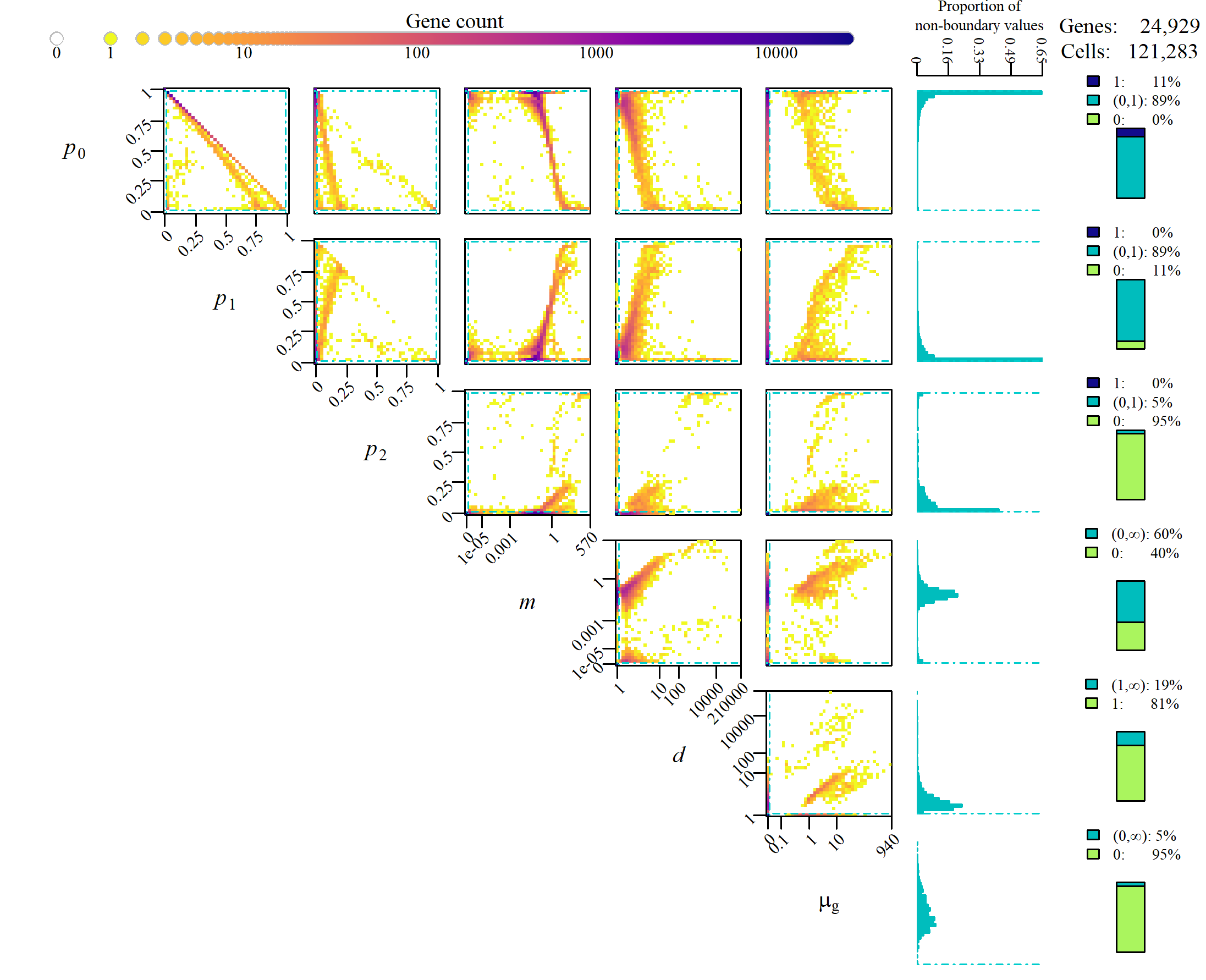}}
\caption{Full set of bivariate and univariate histograms that extends the visualisations of parameter estimates calculated on the CD14 data seen in the article (Section \ref{subsec:res_cd14}). The bivariate histograms plotting $p_0$, $p_1$ and $p_2$ contain the same information as seen in Figure \ref{fig:cd14_tern}, and those plotting $p_0$ against $m$, $d$ and $\mu_g$ are identical to those in Figure \ref{fig:cd14_hists}. The bars to the right show the proportion of genes on each parameter's boundaries. Between these and the bivariate histograms are univariate histograms for the non-boundary values of each parameter. Parameters generally take boundary values as a consequence of model simplification, therefore each component's proportion of boundary values will reflect how frequently they were simplified or omitted.}
\label{fig:cd14_full_hists}
\end{figure*}

\begin{figure*}[!h]%
\centering
{\includegraphics{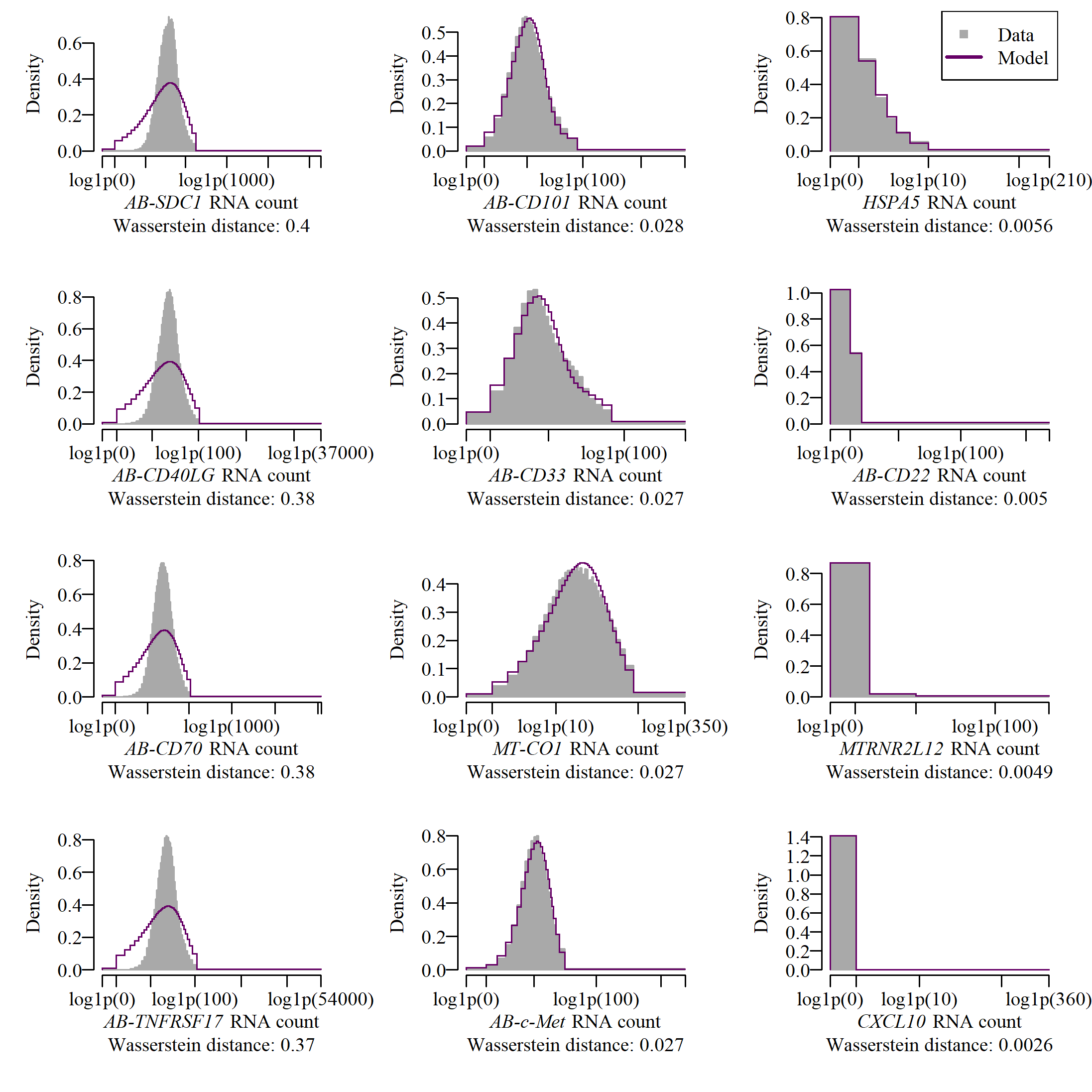}}
\caption{ Histograms of the count data for 12 genes in the CD14 sample with a $p_B$-value of zero. The four in the left column are those with the largest Wasserstein distance, those in the right column are the smallest, and in the centre column are the genes with distances in the middle. The purple lines show the mass functions of the fitted model for that gene.\\
These histograms inform us as to which CD14 genes were poorly fitted, and how a combination of Wasserstein distance and $p_B$-value can be used to filter out problematic genes. The high distance genes with $p_B$-values of zero have the mode of the negative-binomial component in line with the mode of the data, but excessively large estimates of $d$ result in conspicuous differences between the model and empirical distributions. Closer investigation of the model fit at the tail values shows that the geometric component is not able to capture the kurtosis of these genes. As a result, the negative-binomial components are fitted with larger $d$ values to compensate. This lends credence to our argument that negative-binomial models alone are not able to account for the tail values seen in scRNA-seq data. 
The genes with Wasserstein distances near the median have model distributions that appear to capture the overall shape of the data, with the disparities not invalidating the parameter estimates as summary statistics. Similarly, the genes with the smallest Wasserstein distances are well described, with the tail values not appearing to result in a changes in parameter estimates that compromise fit on the majority of count values.}
\label{fig:cd14_lowpb}
\end{figure*}

\begin{figure}[t]%
\centering
{\includegraphics{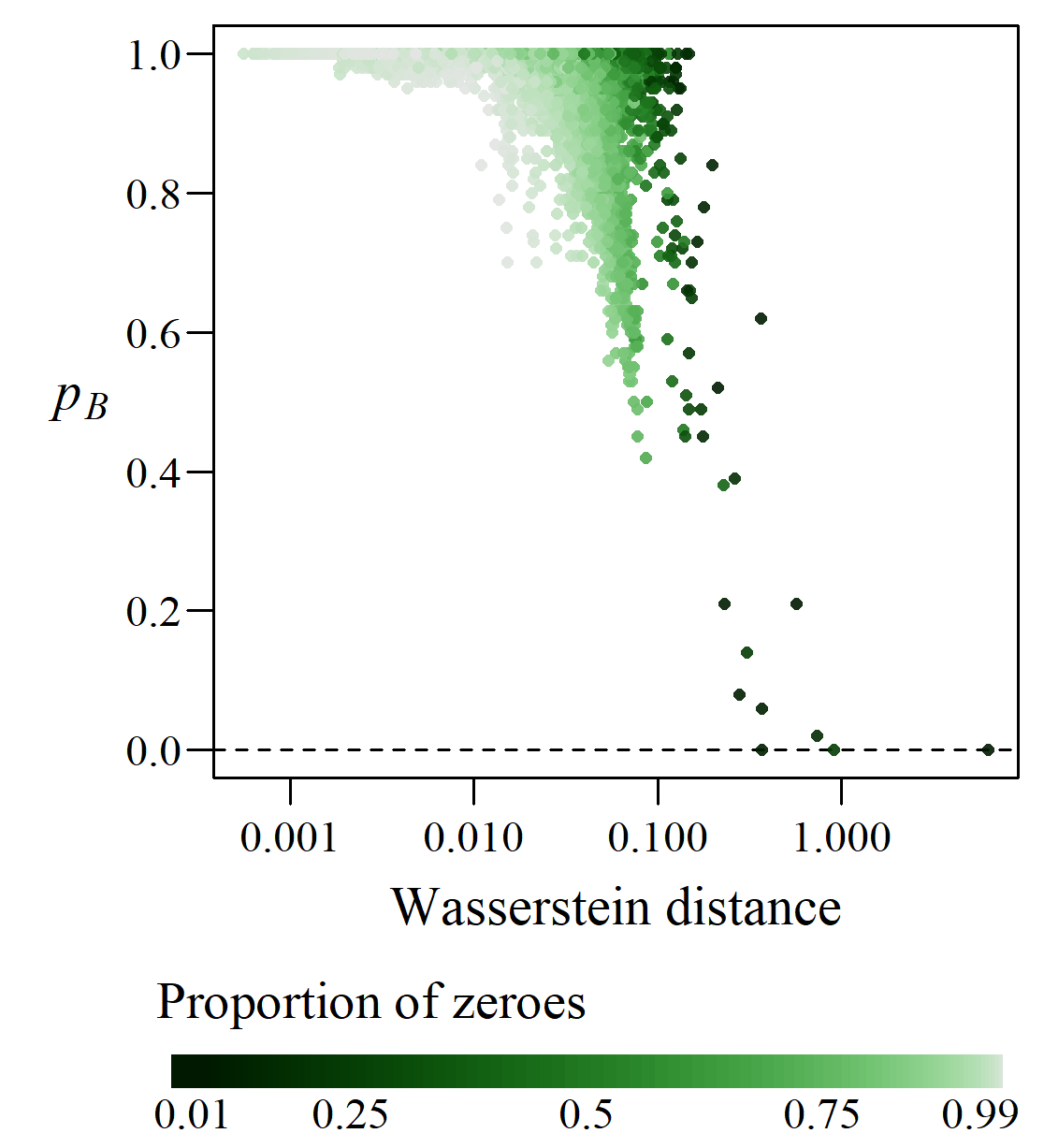}}
\caption{Wasserstein distance and $p_B$-values for the genes in the \textit{T.\ brucei} data \citep{briggs_single-cell_2021}, with points coloured according to the proportion of zero counts for that gene.
The three genes with $p_B$-values of zero have their count data and model distributions visualised in Supplementary Figure \ref{fig:tbruce_pb_zero}. 
An exploration of the six genes with lower, non-zero $p_B$-values revealed that their fitted model distributions closely resembled the empirical distributions. This is consistent with model fitting on this sample being successful, with a single, unambiguous outlier detected.}
\label{fig:tbruce_wass}
\end{figure}

\clearpage
\begin{figure*}[!h]%
\centering
{\includegraphics{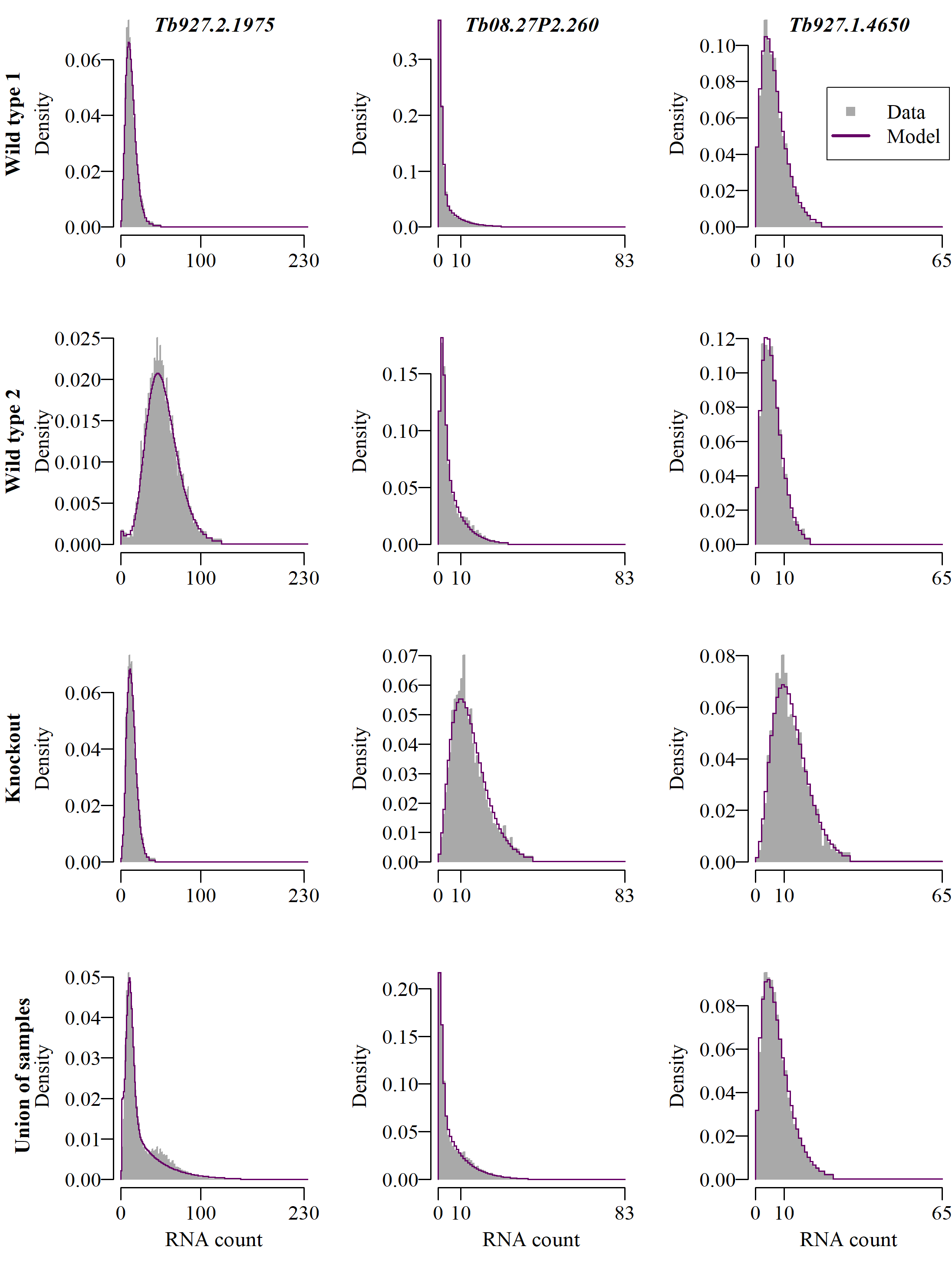}}
\caption{ Extension of Figure \ref{fig:t_bruce_bimod} containing all three genes in the \textit{T. brucei} data \citep{briggs_single-cell_2021} with a $p_B$-value of zero, along with the RNA counts and model densities for the three subsamples.}
\label{fig:tbruce_pb_zero}
\end{figure*}

\clearpage

\section{Supplementary Table}
\label{app:time_table}
\begin{table}[h!]
\begin{center}
\begin{tabular}{ l r r r r }
\hline
& CD14 & \textit{Hierarchicell} & \textit{Muscat} & \textit{T. brucei}\\ 
 \hline
Cell count & 121,283 & 4,000 & 4,000 & 10,894\\
Gene count & 24,929  & 2,000 & 2,000 & 9,066\\
Computation time without diagnostics & \texttt{1h.04m.} & \texttt{2m.} & \texttt{2m.} & \texttt{10m.}\\
Computation time with diagnostics & \texttt{5h.12m.}& \texttt{3m.} & \texttt{6m.} & \texttt{29m.}\\
\hline
\end{tabular}
\end{center}
		\caption{Table containing the number of cells and genes, as well as computation times with and without Wasserstein diagnostics, for the four datasets used in this paper. All analyses were performed on a laptop with an Intel i5-1135G7 CPU at 2.40 GHz with 16GB of RAM. The speed of these calculations could be further improved by taking advantage of parallelisation.}
\label{table:comp_times}
\end{table}

\end{document}